# Rapid host switching in generalist *Campylobacter* strains erodes the signal for tracing human infections

Bethany L. Dearlove[1§], Alison J. Cody[2], Ben Pascoe[3,4], Guillaume Méric[3,4], Daniel J. Wilson,[1,5]*, Samuel K. Sheppard[2,3,4]*

[1]Nuffield Department of Medicine, Experimental Medicine Division, University of Oxford, Oxford, UK

[2]Department of Zoology, University of Oxford, Oxford, UK

[3]College of Medicine, Institute of Life Science, Swansea University, Swansea, UK

[4]MRC CLIMB Consortium, Institute of Life Science, Swansea University, Swansea, UK

[5]Wellcome Trust Centre for Human Genetics, University of Oxford, Oxford, UK

*These authors contributed equally to this work.

§Corresponding author. Present address: Department of Veterinary Medicine, University of Cambridge, Cambridge, United Kingdom. Email: bd357@cam.ac.uk

**Running Title:** Rapid zoonosis in generalist *Campylobacter* strains

**Subject Category:** Evolutionary genetics

# Abstract

*Campylobacter jejuni* and *Campylobacter coli* are the biggest causes of bacterial gastroenteritis in the developed world, with human infections typically arising from zoonotic transmission associated with infected meat, especially poultry. Because this organism is not thought to survive well outside of the gut, host associated populations are genetically isolated to varying degrees. Therefore the likely origin of most *Campylobacter* strains can be determined by host-associated variation in the genome. This is instructive for characterizing the source of human infection at the population level. However, some common strains appear to have broad host ranges, hindering source attribution. Whole genome sequencing has the potential to reveal fine-scale genetic structure associated with host specificity within each of these strains.

We found that rates of zoonotic transmission among animal host species in ST-21, ST-45 and ST-828 clonal complexes were so high that the signal of host association is all but obliterated. We attributed 89% of clinical cases to a chicken source, 10% to cattle and 1% to pig. Our results reveal that common strains of *C. jejuni* and *C. coli* infectious to humans are adapted to a generalist lifestyle, permitting rapid transmission between different hosts. Furthermore, they show that the weak signal of host association within these complexes presents a challenge for pinpointing the source of clinical infections, underlining the view that whole genome sequencing, powerful though it is, cannot substitute for intensive sampling of suspected transmission reservoirs.

# Introduction

*Campylobacter jejuni* and *Campylobacter coli* are zoonotic pathogens with broad host ranges, carried, apparently asymptomatically, as part of the gut microbiota of a range of wild and domesticated mammal and bird species. Common carriage in the gut of animals and poultry farmed for meat leads to numerous opportunities for contamination of food products, which may take place at various points from farm to fork (Neimann *et al.*, 2003). Human-to-human transmission is rare (Allos, 2001), hence *Campylobacter* infection in humans tends to be sporadic, and seldom manifests as outbreaks except when a single point source results in direct transmission to many people, for example via contaminated drinking water (Palmer *et al.*, 1983; Pebody *et al.*, 1997; Engberg *et al.*, 1998; Clark *et al.*, 2003).

The sources of human *Campylobacter* infection have been well characterized at the population level. Genetic analysis, particularly using seven-locus multilocus sequence typing (MLST) data, has helped to attribute the sources of clinical infections by exploiting differences in the frequency of *Campylobacter* strains that live in different animal and environmental reservoirs (McCarthy *et al.*, 2007). For example, isolates belonging to related sequence types (STs) from the ST-257 and ST-61 clonal complexes are strongly associated with chicken and ruminants respectively, while wild bird species are often colonised by phylogenetically divergent *Campylobacter* lineages (Sheppard *et al.*, 2011; Griekspoor *et al.*, 2013). Independent MLST-based studies in England, Scotland and New Zealand found that over 95% of human infections are attributable to animals farmed for meat and poultry, with 56-76% of cases attributable to poultry in particular (Wilson *et al.*, 2008; Mullner *et al.*, 2009; Sheppard *et al.*, 2009).

However, at the individual isolate level, there is often considerable uncertainty in source attribution because some of the most common disease causing strains, notably those belonging to the ST-21, ST-45 and ST-828 clonal complexes, are regularly isolated from multiple animal species. This means that quantitative attribution to a single host reservoir is difficult using MLST data alone. The reason for the apparently broader host range of these lineages is not currently understood, and may reflect true host generalism (Gripp *et al.*, 2011; Sheppard *et al.*, 2014) or the existence of host restricted sublineages within these clonal complexes.

The increasing availability of whole genome sequence data provides a potential solution to the challenge of attributing the origin of generalist strains in clinical samples, as a host-specific signal might be gleaned from the remaining 99.8% of the genome outside the seven loci sequenced by standard MLST. Genomics has been used to investigate direct transmission events in a range of bacterial species including *Staphylococcus aureus* (Harris *et al.*, 2010; Price *et al.*, 2014), *Clostridium difficile* (Eyre *et al.*, 2013) and *Mycobacterium tuberculosis* (Walker *et al.*, 2013). This raises the prospect of enhancing understanding of *Campylobacter* epidemiology, for example by identifying individual retailers or producers of *Campylobacter*-contaminated food in cases of human infection, and detecting cryptic point source outbreaks (Cody *et al.*, 2013). Ultimately this information can help to inform targeted interventions and reduce the incidence of human campylobacteriosis (Sears *et al.*, 2011).

In this study we investigated the utility of whole genome sequence data for improving the accuracy with which human cases can be attributed to particular animal reservoirs, by focusing on common strains that are difficult to attribute with MLST. Specifically, we sought to detect fine-scale host-specific genetic structuring within *C. jejuni* ST-21 and ST-45 complex and *C. coli* ST-828 complex isolates sampled from different animal species and to exploit that signal to improve human source attribution. We observed extraordinarily rapid rates of zoonotic transfer within these strains leading to little or no phylogenetic association with host species, limiting the ability of whole genome sequencing to improve source attribution, but revealing new insight into the transmission dynamics within these common *Campylobacter* strains.

## Materials and Methods

### Sequences

The ST-21 and ST-45 (*C. jejuni*) and ST-828 (*C. coli*) clonal complexes were chosen as the focus of the study because they are among the most common lineages causing human disease, but are difficult to attribute to source populations using seven-locus MLST data (Wilson *et al.*, 2008; Sheppard *et al.*, 2009). Isolates from these three clonal complexes were chosen from MLST collections (www.pubmlst.org/campylobacter, Jolley and Maiden (2010)), sequenced and assembled according to the protocols described in Sheppard et al. (2013) and Cody et al. (2013). In total, 348, 87 and 158 sequences were obtained for the ST-21, ST-45 and ST-828 complexes, respectively (Lefébure *et al.*, 2010; Cody *et al.*, 2013; Sheppard, Didelot, Jolley, *et al.*, 2013; Sheppard, Didelot, Meric, *et al.*, 2013; Sheppard *et al.*, 2014) (see Supplementary Table 1). Isolates belonging these complexes are predominantly associated with chicken and cattle, from where the majority of animal samples were obtained (13 chicken isolates from each of clonal complexes, and 7, 9 and 10 cattle isolates for the ST-21, ST-45 and ST-828 clonal complexes respectively). In addition, three ST-45 complex isolates were included from wild bird species, and six ST-828 complex isolates from pigs. Clinical disease isolates (328, 68 and 128 for the ST-21, ST-45 and ST-828 clonal complexes respectively) were obtained from a surveillance study of samples submitted to the microbiology laboratory of the John Radcliffe Hospital, Oxfordshire, United Kingdom (UK).

After Illumina sequencing, the high coverage short reads were *de novo* assembled using Velvet (Zerbino and Birney, 2008), and the resulting contiguous sequences ('contigs') stored using BIGSdb (Jolley and Maiden, 2010). These contigs were then compared to the NCTC11168 reference sequence (Genbank accession number: AL111168) to identify genes using a BLAST search (Parkhill *et al.*, 2000; Gundogdu *et al.*, 2007). Orthologous genes were defined as homologous genes that had 70% or greater nucleotide identity, and less than 50% difference in alignment length. Genes for all isolates were then aligned using MUSCLE (Edgar, 2004), and concatenated into a single sequence per sample.

**Accounting for ancestral recombination**

There is good evidence that novel diversity in *Campylobacter* is generated frequently by the continued movement of genes between lineages, more so even than by the evolution of new variants through mutation (Wilson *et al.*, 2009; Sheppard *et al.*, 2010). High levels of recombination lead to mosaic genomes with differing ancestral histories, and the full evolution of samples cannot be represented by a coalescent genealogy.

During preliminary analyses using BEAST (Drummond *et al.*, 2012), a relaxed clock model and gamma site heterogeneity were used to try and account for the effect of recombination. However, these analyses diagnosed difficulties in the mixing of the MCMC algorithm underlying BEAST, with multiple runs of the same analysis frequently converging to different topologies. To overcome this, and to account for the effect that recombination has in skewing the branch lengths of the dominant phylogenetic tree (Schierup and Hein, 2000), homoplasious sites incompatible with the maximum likelihood tree were identified and removed (Pupko *et al.*, 2000; Guindon *et al.*, 2010). For the final dataset, only the biallelic sites compatible with the inferred phylogeny were included in the alignment, alongside all of the non-variable sites. The removal of homoplasies in this way resolved the mixing issues with BEAST. In addition, to improve computation time for later analyses, missing alleles were imputed using ClonalFrameML (Pupko *et al.*, 2000; Didelot and Wilson, 2015).

**Phylogenetic analysis**

The analysis was implemented in BEAST v1.7.5 (Drummond *et al.*, 2012), assuming a constant population size. All parameters were scaled in terms of the effective population size, $N_e$, by fixing it to 1.0. The HKY85 model of nucleotide substitution (Hasegawa *et al.*, 1985) was used with an uncorrelated log-normal relaxed clock and gamma rate heterogeneity with four categories (Drummond *et al.*, 2006). A log-normal prior with a mean of 1.0 and standard deviation of 1.25 on the logarithmic-scale was assumed for the transition:tranversion ratio $\kappa$, and an exponential prior with a mean of 1.0 was utilized for the gamma shape parameter $\alpha$. For the log-normal relaxed clock parameters, a uniform prior between 0.0 and 10.0 was assumed for the mean, and an exponential with mean 1.0 for the standard deviation. A uniform (Dirichlet) prior was used for the nucleotide frequencies.

We used the phylogeographic model by Lemey et al. (Lemey *et al.*, 2009) to reconstruct zoonotic transmission and infer the source of the infection (host species) along the branches. This allows us to report the posterior probability of any node in the tree being a particular state (i.e. the most likely host of the ancestral lineage), and also to obtain relative rates of transition between states (i.e. switching between host species). Given that human to human transmission is rare (Allos, 2001), an ambiguity code was set up in the zoonotic model to allow the human isolates to have an 'unknown' source population. For each human isolate, the other host species in the analysis were given equal prior probability and thus most likely source of the human isolates could be inferred. A uniform prior from 0.0 to 100.0 was assumed for the host migration rate, a gamma prior with mean 1.0 and scale 1.0 for the relative rates of migration, and a uniform (Dirichlet) prior for the host population equilibrium frequencies. To calculate the number of discrete zoonotic transmissions across the branches, Markov jumping was performed using BEAST (Minin and Suchard, 2008a, 2008b; Talbi *et al.*, 2010).

The MCMC was run for 500 million iterations, with samples taken every 10,000 iterations. For each ST, the analysis was repeated twice (three times for ST-21) with different initial values to check convergence and mixing, and these runs were combined for the final results. Unless otherwise stated, the posterior median was used for point estimates and the (2.5%, 97.5%) quantiles for credible intervals. The inferred ancestral host type for each branch in the phylogeny was taken to be the one with the highest posterior probability. Sample dates were not available for all isolates, so the analysis is given in coalescent time (denoted $\tau$) and calibrated to years using an independent estimate of the mutation rate in *Campylobacter* of $3.23 \times 10^{-5}$ substitutions per site per year from Wilson et al. (2009).

## Results

**Fine-scale Phylogenetic Structure within Sequence Types**

Within the ST-21, ST-45 and ST-828 complexes, isolates from different host species were often more closely related than those isolated from the same host species (Figure 1). If the strains harboured previously undetected sub-ST lineages that were strongly host associated, one would expect to see distinct clusters of the same coloured branches together. However, this was not the case, with branches of the same colour – representing the reconstructed reservoir population of that lineage – scattered throughout the tree in all STs. Furthermore, one could expect that isolates from mammalian hosts would be more closely related than those from avian species reflecting enhanced transfer potential between physiologically similar hosts. Again, this was not observed, with bird isolates in the ST-45 complex and pig isolates in ST-828 complex being closely related to both chicken and cattle isolates. This may suggest that the ability to colonise a specific host has either evolved several times throughout the tree, most plausibly through horizontal gene transfer given that mutation is rare, or that the isolates are adapted to infect all species of host in the sample.

Much of the ancestral history of the lineages was inferred to have occurred within the chicken host population, shown by the dominance of yellow branches. The most recent common ancestor (MRCA) of all three clonal complexes was inferred to have colonized chicken, with posterior probabilities of 0.612, 0.498 and 0.500 for the ST-21, ST-45 and ST-828 complexes respectively.

**Rates of Zoonotic Transmission in *Campylobacter***

Under the host restricted hypothesis, isolates sampled from the same source reservoir will cluster into a single clade within the phylogeny. Since we sampled ST-21 complex isolates from two host species (chicken and cattle), there must have been exactly one zoonotic transfer in the tree under the restricted host hypothesis, or more under the generalist model. We sampled ST-45 and ST-828 complex isolates from three host species (chicken, cattle and wild birds in ST-45 or swine in ST-828), necessitating exactly two zoonotic transfer events in the tree under the restricted host hypothesis. In fact, the total number of migration events for all three clonal complexes was estimated to be much higher than these minimum values, with 588.9 (95% credibility interval: 109.8, 1325.9), 468.7 (105.7, 1264.5) and 117.7 (36.6, 456.3) cross-species transmission events for ST-21, ST-45 and ST-828 complexes respectively. In all three cases, the number of zoonotic transfers under the restricted host hypothesis was outside the range of the 95% credible intervals, demonstrating that isolates from these clonal complexes display host generalism and excluding the possibility of host restricted sub-lineages.

The overall estimated migration rate was much more frequent in the *C. jejuni* clonal complexes compared to the *C. coli* (ST-828) complex. Using a mutation rate of $3.23 \times 10^{-5}$ substitutions per site per year for calibration (Wilson *et al.*, 2009), this corresponds to approximately one host jump every twelve years for ST-828 compared to one every 1.6 or 1.8 years in the ST-21 and ST-45 complexes. Although the credible intervals are wide, particularly for the *C. jejuni* sequence types, the finding of extraordinarily rapid rates of zoonotic transfer were robust to alternative prior beliefs. Thus, based upon the isolate collection in this study, the estimated rate of zoonosis is sufficiently rapid within ST-21 and ST-45 clonal complexes that there is no association between genetic structure and host species. This is consistent with a generalist lifestyle in which the isolates are equally adapted to transmission between versus within species.

**Tracing the Source of Human Infection**

In Figure 1, the human cases at the tips of the tree are represented with black circles, and the posterior probability of the sources for the human isolates are illustrated by the bar plots. The majority of clinical cases (462 out of 519 cases, or 89%) were attributed to a chicken source, with 53 cases (10%) attributed to cattle and 4 (1%) to pig (Figure 2). This is consistent with results found by MLST (Wilson *et al.*, 2008; Mullner *et al.*, 2009; Sheppard *et al.*, 2009), but the difficulty of attributing individual human cases to specific source species with high confidence, even using whole genomes, reflects the remarkable transmission rates of lineages among host species. Even when human isolates were most closely related to clades sampled only from chicken, there remained appreciable probabilities (in the range 30-40%) that the direct source of transmission was non-chicken. This degree of uncertainty is comparable to the 67% accuracy of source attribution that Wilson et al. (2008) achieve with MLST alone.

**Discussion**

We investigated the source of *Campylobacter* infection in humans using whole genome sequencing, focusing on STs that are frequently isolated from multiple host species and are therefore weakly host-associated on the basis of MLST (Wilson *et al.*, 2008; Sheppard *et al.*, 2009, 2014; Gripp *et al.*, 2011). We tested whether these STs were aggregations of strongly host-restricted sub-lineages, or whether they represented genuine generalists (Gripp *et al.*, 2011). We estimated rates of zoonotic transfer between *Campylobacter* reservoir species, and attributed individual clinical cases to animal sources. Consistent with previous studies (Harris *et al.*, 1986; Wingstrand *et al.*, 2006; Wilson *et al.*, 2008; Mullner *et al.*, 2009; Sheppard *et al.*, 2009), we identified the chicken reservoir as accounting for the majority of human *Campylobacter* infections, emphasizing the importance of measures aimed at controlling food-borne disease in agriculture and the food industry.

We found that fine-scale population structure across the genome within the ST-21, ST-45 and ST-828 clonal complex isolates was not host-associated. This is consistent with the existence of genuine generalist strains, adapted to transmit between and live within multiple host species. There are clear advantages of a generalist lifestyle in agricultural animal species where multiple mammalian and avian species routinely live in close proximity, giving rise to frequent opportunity for zoonotic transmission.

Some differences were observed in the relative rate of host switching between *C. coli* and *C. jejuni*. Specifically, for the *C. coli* ST-828 complex isolates, there was evidence for slower rates of zoonotic transfer, compared to the *C. jejuni* ST-21 and ST-45 complex isolates. We estimated that there were 588 migration events across the tree in the ST-21 complex and 468 in the ST-45 complex, compared to only 117 in ST-828. In all cases we were able to reject the hypothesis of a unique host jump founding the population in each new species, with significantly more migration events than would be expected if isolates were host restricted (Table 2). This is evidenced by the scattering of isolates sampled from different sources throughout the phylogeny.

The overall rates of migration between different host species provides information about the ecology of the generalist *Campylobacter* strains. Within the ST-45 complex, there was very little difference in the relative rates of transmission between host species (Table 2). However, in ST-828 complex isolates, host switches were twice as frequent between cattle and swine compared with the rates of migration between these species and chicken. This disparity between mammal-mammal and mammal-bird transmission rates suggests that the efficacy of zoonotic transmission may be lower in the ST-828 complex compared to ST-21 or ST-45 complexes. Mammals and birds exhibit many physiological differences including a difference in core body temperature of about 38°C in cattle and pigs versus 42°C in chickens. Variation in rates of zoonosis may be explained by differences in the route of transmission between the mammalian species, either through opportunity – for example if cattle and pigs have more commonly shared the same physical environment, or through affinity – involving factors associated with the ability to colonize different species. Suggestively, the average genome size of ST-828 complex isolates is smaller than that of ST-21 and ST-45 complex isolates, potentially reflecting more limited phenotypic plasticity that would limit the ability to occupy multiple divergent niches.

By comparing an approach using relatively few isolates characterized at many loci, by whole genome sequencing, with methods using hundreds or thousands of isolates characterized at few loci, such as MLST studies (Wilson *et al.*, 2008; Mullner *et al.*, 2009; Sheppard *et al.*, 2009), we investigated the added benefit of whole genome sequencing for investigating lineages with rapid host-switching within clonal complexes. There was little additional information regarding source of infection provided by the whole genome sequence in the majority of human cases. This suggests an inherent trade-off between the number of samples and the number of loci sequenced, and that the approximately thousand-fold increase in sequence information afforded by whole genomes over 7-locus MLST does not, on its own, eliminate the need for detailed sampling. While there is evidence of the potential for whole genome sequencing to provide greater resolution in detecting recent transmission of human *Campylobacter* infection from non-chicken sources, this study highlights the importance of intensive sampling of potential source populations and an understanding of host transmission ecology for effective source attribution.

## Acknowledgements

This study was supported by the Oxford NIHR Biomedical Research Centre and the UKCRC Modernising Medical Microbiology Consortium, the latter funded under the UKCRC Translational Infection Research Initiative supported by the Medical Research Council, the Biotechnology and Biological Sciences Research Council and the National Institute for Health Research on behalf of the UK Department of Health (Grant G0800778) and the Wellcome Trust (Grant 087646/Z/08/Z). BLD is supported by a Medical Research Council Methodology Research Programme grant (grant number MR/J013862/1). AJC was supported by the United Kingdom Department for Environment, Food, and Rural Affairs and Food Standards Agency (grant number OZ0624). DJW is a Sir Henry Dale Fellow, jointly funded by the Wellcome Trust and the Royal Society (Grant 101237/Z/13/Z). SKS is funded by the Biotechnology and Biological Sciences Research Council (BBSRC), the Medical Research Council (MR/L015080/1) and the Wellcome Trust. This publication made use of the Campylobacter Multi Locus Sequence Typing website (http://pubmlst.org/campylobacter/) developed by Keith Jolley and sited at the University of Oxford (24). The development of this site has been funded by the Wellcome Trust.

Supplementary information is available at The ISME Journal's website.

## Conflicts of Interest

The authors have no conflicts of interest to declare.

# Figures and Tables

**Figure 1: Maximum clade credibility trees for the ST-21, ST-45 ST-828 complexes.**

Tips are coloured by host from which the sample was isolated: chicken (yellow), cattle (red), pig (pink), wild bird (green) and human (black). Branches are coloured according to the ancestral source population inferred using the maximum posterior probability. Pie charts show the posterior probability for the root of the tree. For each human case, the posterior probability of source is shown as a stacked bar plot. Scale is given in units of coalescent time. Note that a change in host may have occurred at any point on the branch, not necessarily at the node, and it is also possible to have a number of host switches occurring along a branch.

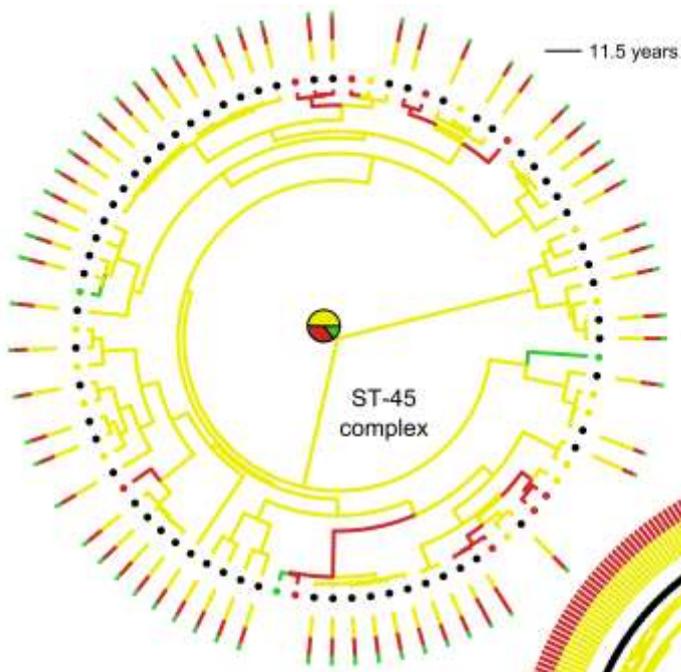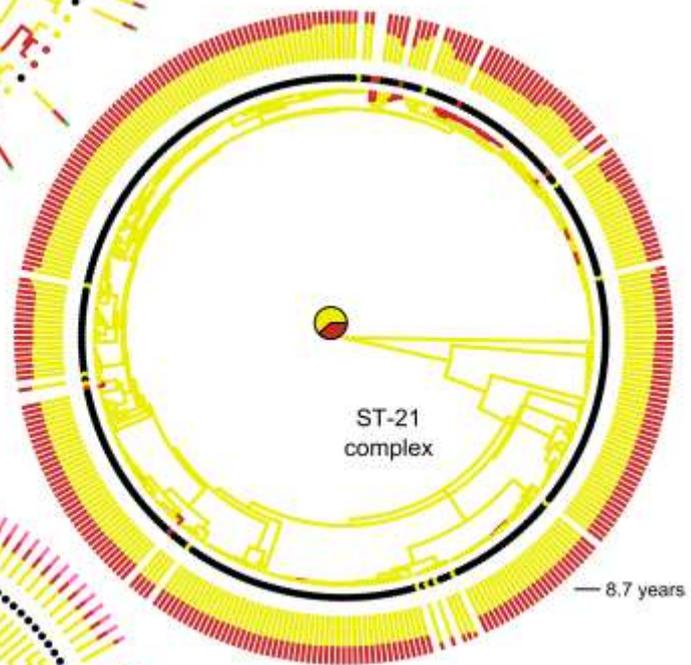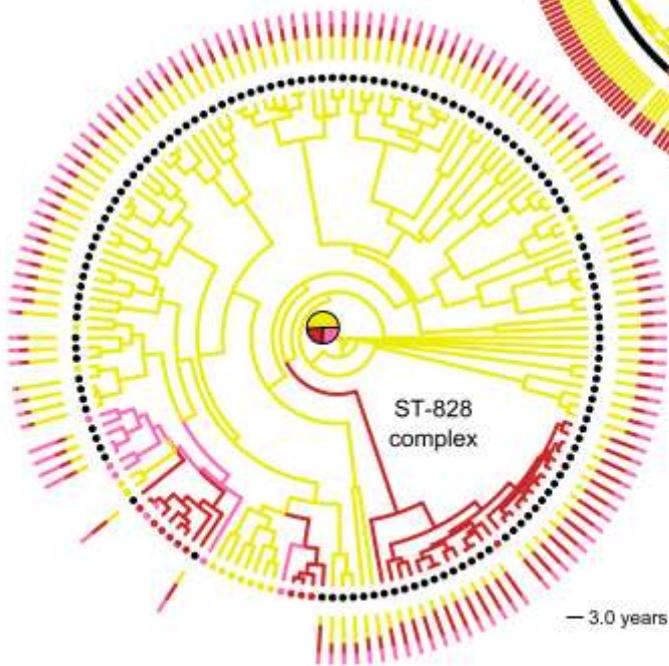

**Figure 2: Probability of source for clinical cases for a) ST-21, b) ST-45 and c) ST-828.**

The posterior probability of each human isolate (vertical bars) broken down by source population: chicken (yellow), cattle (red), pig (pink) and wild bird (green). The isolates have been reordered along the x-axis for visualisation purposes.

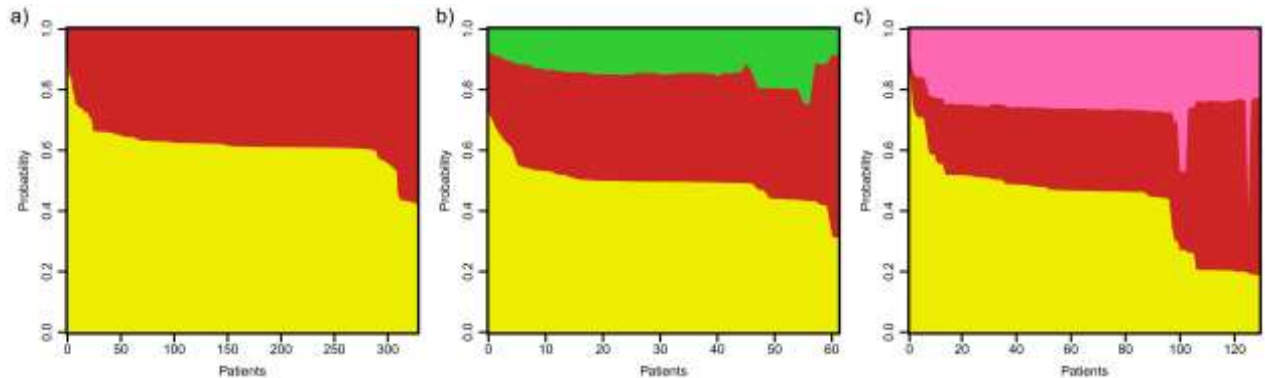

**Table 1: Summary of sequence data for the three STs.**

|  | **ST-21 complex** | **ST-45 complex** | **ST-828 complex** |
|---|---|---|---|
| **Genome alignment length:** | **1544595** | **1465323** | **1421603** |
| Non-polymorphic sites | 1447536 | 1386349 | 1272113 |
| Polymorphic sites | 97059 | 78929 | 149490 |
| **Biallelic sites:** | **90239** | **722233** | **136375** |
| Compatible with ML tree | 41895 | 34960 | 60692 |
| **Total sites used in analysis** | **1489431** | **1421309** | **1332805** |

**Table 2: Parameter estimates for each ST.**

| Parameter | ST-21 complex | ST-45 complex | ST-828 complex |
|---|---|---|---|
| **Substitution rate** | | | |
| ($10^{-3}$ /site /$\tau$) | 2.815 (1.950, 3.560) | 3.715 (2.315, 5.158) | 9.582 (8.019, 11.534) |
| **TMRCA** | | | |
| Coalescent time, $\tau$ | 1.022 (0.442, 3.068) | 0.709 (0.288, 2.583) | 0.159 (0.110, 0.232) |
| Years[1] | 89.069 (38.521, 267.381) | 81.545 (33.124, 297.085) | 47.168 (32.632, 68.824) |
| **Host switching rate** | | | |
| /$\tau$ | 54.078 (10.089, 97.519) | 61.634 (14.540, 98.069) | 23.491 (6.802, 85.357) |
| /year[1] | 0.621 (0.116, 1.119) | 0.536 (0.126, 0.853) | 0.079 (0.023, 0.287) |
| **Number of migrations** | | | |
| | 588.89 (109.80, 1325.90) | 468.67 (105.70, 1264.48) | 117.691 (36.558, 456.318) |
| **Estimated host frequencies** | | | |
| Chicken | 0.614 (0.388, 0.809) | 0.498 (0.311, 0.685) | 0.425 (0.136, 0.691) |
| Cattle | 0.386 (0.191, 0.612) | 0.348 (0.185, 0.540) | 0.282 (0.095, 0.562) |
| Wild bird | - | 0.140 (0.0429, 0.313) | - |
| Pig | - | - | 0.270 (0.103, 0.545) |
| **Relative migration rates between host species** | | | |
| Chicken-Cattle | 0.695 (0.023, 3.705) | 0.820 (0.051, 3.779) | 0.303 (0.010, 2.139) |
| Chicken-Wild Bird | - | 0.592 (0.021, 3.282) | - |
| Chicken-Pig | - | - | 0.581 (0.024, 3.129) |
| Cattle-Wild Bird | - | 0.728 (0.031, 3.643) | - |
| Cattle-Pig | - | - | 1.367 (0.192, 4.783) |

[1]Using the mutation rate $3.23 \times 10^{-5}$ substitutions per site per year [21] for calibration with the median evolutionary rate from the BEAST analysis, one unit of coalescent time ($\tau$) is equal to: 87.151 years in the ST-21 complex, 115.015 years in ST-45 complex, and 296.656 years in ST-828 complex. The median value was used for estimating all other parameters in years.

**Supplementary Table 1: Isolates included in the study.**

All isolates are available from http://pubmlst.org/campylobacter/.

*Sequence type was derived from the allelic profile of seven housekeeping genes by multilocus sequence typing (MLST).

**Clonal complex is defined as including any ST that matches a previously defined central genotype (http://pubmlst.org/campylobacter/) at three or more loci.

***References refer to:

| ID | Isolate | source | species | aspA | glnA | gltA | glyA | pgm | tkt | uncA | Sequence Type* | Clonal Complex** | Reference*** |
|---|---|---|---|---|---|---|---|---|---|---|---|---|---|
| 1 | NC_002163 | human | *Campylobacter jejuni* | 2 | 1 | 5 | 3 | 4 | 1 | 5 | 43 | ST-21 complex | 2 |
| 2 | CampsClin262 | human | *Campylobacter jejuni* | 2 | 1 | 1 | 3 | 2 | 1 | 3 | 262 | ST-21 complex | 2 |
| 3 | CampsClin266 | human | *Campylobacter jejuni* | 2 | 1 | 5 | 3 | 2 | 54 | 5 | 266 | ST-21 complex | 2 |
| 4 | CampsClin883 | human | *Campylobacter jejuni* | 2 | 17 | 2 | 3 | 2 | 1 | 5 | 883 | ST-21 complex | 2 |
| 5 | chicka21 | chicken | *Campylobacter jejuni* | 2 | 1 | 1 | 3 | 2 | 1 | 5 | 21 | ST-21 complex | 2 |
| 6 | cow518 | cattle | *Campylobacter jejuni* | 2 | 1 | 1 | 3 | 2 | 1 | 5 | 21 | ST-21 complex | 2 |
| 7 | CampsClin53 | human | *Campylobacter jejuni* | 2 | 1 | 21 | 3 | 2 | 1 | 5 | 53 | ST-21 complex | 2 |
| 8 | cowa21 | cattle | *Campylobacter jejuni* | 2 | 1 | 1 | 3 | 2 | 1 | 5 | 21 | ST-21 complex | 2 |
| 9 | chickc21 | chicken | *Campylobacter jejuni* | 2 | 1 | 1 | 3 | 2 | 1 | 5 | 21 | ST-21 complex | 2 |
| 10 | chick104 | chicken | *Campylobacter jejuni* | 2 | 1 | 1 | 3 | 7 | 1 | 5 | 104 | ST-21 complex | 2 |
| 11 | chick19 | chicken | *Campylobacter jejuni* | 2 | 1 | 12 | 3 | 2 | 1 | 5 | 50 | ST-21 complex | 2 |
| 12 | chick50 | chicken | *Campylobacter jejuni* | 2 | 1 | 12 | 3 | 2 | 1 | 5 | 50 | ST-21 complex | 2 |
| 13 | chick53 | chicken | *Campylobacter jejuni* | 2 | 1 | 21 | 3 | 2 | 1 | 5 | 53 | ST-21 complex | 2 |
| 14 | chick262 | chicken | *Campylobacter jejuni* | 2 | 1 | 1 | 3 | 2 | 1 | 3 | 262 | ST-21 complex | 2 |
| 15 | chick266 | chicken | *Campylobacter jejuni* | 2 | 1 | 5 | 3 | 2 | 54 | 5 | 266 | ST-21 complex | 2 |
| 16 | chick1086 | chicken | *Campylobacter jejuni* | 2 | 1 | 12 | 3 | 2 | 1 | 5 | 50 | ST-21 complex | 2 |
| 17 | chick1360 | chicken | *Campylobacter jejuni* | 2 | 1 | 12 | 3 | 2 | 1 | 5 | 50 | ST-21 complex | 2 |
| 18 | cowb21 | cattle | *Campylobacter jejuni* | 2 | 1 | 1 | 3 | 2 | 1 | 5 | 21 | ST-21 complex | 2 |
| 19 | cow104 | cattle | *Campylobacter jejuni* | 2 | 1 | 1 | 3 | 7 | 1 | 5 | 104 | ST-21 complex | 2 |
| 20 | cow3201 | cattle | *Campylobacter jejuni* | 2 | 1 | 5 | 3 | 2 | 1 | 5 | 19 | ST-21 complex | 2 |
| 21 | chickb21 | chicken | *Campylobacter jejuni* | 2 | 1 | 1 | 3 | 2 | 1 | 5 | 21 | ST-21 complex | 3 |
| 22 | chick883 | chicken | *Campylobacter jejuni* | 2 | 17 | 2 | 3 | 2 | 1 | 5 | 883 | ST-21 complex | 3 |
| 23 | CampsClin21 | human | *Campylobacter jejuni* | 2 | 1 | 1 | 462 | 2 | 1 | 5 |  | ST-21 complex | 3 |
| 24 | OxClina21 | human | *Campylobacter jejuni* | 2 | 1 | 1 | 3 | 2 | 1 | 5 | 21 | ST-21 complex | 3 |
| 25 | CjLMG9879 | human | *Campylobacter jejuni* | 2 | 1 | 1 | 5 | 2 | 1 | 5 | 47 | ST-21 complex | 4 |
| 26 | Cj2008-1025 | human | *Campylobacter jejuni* | 2 | 1 | 12 | 3 | 2 | 1 | 5 | 50 | ST-21 complex | 4 |
| 27 | Cj2008-831 | human | *Campylobacter jejuni* | 2 | 1 | 12 | 3 | 2 | 1 | 5 | 50 | ST-21 complex | 4 |
| 28 | Cj110-21 | cattle | *Campylobacter jejuni* | 2 | 1 | 2 | 3 | 2 | 1 | 5 | 982 | ST-21 complex | 4 |

| | | | | | | | | | | | | |
|---|---|---|---|---|---|---|---|---|---|---|---|---|
| 29 | Cj87330 | chicken | *Campylobacter jejuni* | 2 | 1 | 12 | 3 | 2 | 1 | 5 | 50 | ST-21 complex | 4 |
| 30 | Cj1928 | cattle | *Campylobacter jejuni* | 2 | 1 | 1 | 3 | 140 | 3 | 5 | 806 | ST-21 complex | 4 |
| 31 | OXC5333 | human | *Campylobacter jejuni* | 2 | 1 | 12 | 3 | 2 | 1 | 5 | 50 | ST-21 complex | 5 |
| 32 | OXC5335 | human | *Campylobacter jejuni* | 2 | 1 | 12 | 3 | 2 | 1 | 5 | 50 | ST-21 complex | 5 |
| 33 | OXC5344 | human | *Campylobacter jejuni* | 2 | 1 | 1 | 3 | 2 | 1 | 5 | 21 | ST-21 complex | 5 |
| 34 | OXC5349 | human | *Campylobacter jejuni* | 2 | 1 | 12 | 3 | 2 | 1 | 5 | 50 | ST-21 complex | 5 |
| 35 | OXC5350 | human | *Campylobacter jejuni* | 2 | 1 | 1 | 3 | 2 | 1 | 5 | 21 | ST-21 complex | 5 |
| 36 | OXC5364 | human | *Campylobacter jejuni* | 2 | 1 | 1 | 3 | 2 | 1 | 5 | 21 | ST-21 complex | 5 |
| 37 | OXC5372 | human | *Campylobacter jejuni* | 2 | 1 | 5 | 3 | 2 | 1 | 5 | 19 | ST-21 complex | 5 |
| 38 | OXC5378 | human | *Campylobacter jejuni* | 2 | 1 | 21 | 3 | 2 | 1 | 5 | 53 | ST-21 complex | 5 |
| 39 | OXC5393 | human | *Campylobacter jejuni* | 2 | 1 | 12 | 3 | 2 | 1 | 5 | 50 | ST-21 complex | 5 |
| 40 | OXC5397 | human | *Campylobacter jejuni* | 2 | 1 | 1 | 5 | 11 | 343 | 5 | 5242 | ST-21 complex | 5 |
| 41 | OXC5474 | human | *Campylobacter jejuni* | 2 | 1 | 1 | 3 | 2 | 1 | 5 | 21 | ST-21 complex | 5 |
| 42 | OXC5476 | human | *Campylobacter jejuni* | 2 | 1 | 5 | 3 | 2 | 1 | 5 | 19 | ST-21 complex | 5 |
| 43 | OXC5664 | human | *Campylobacter jejuni* | 2 | 1 | 12 | 3 | 2 | 1 | 5 | 50 | ST-21 complex | 5 |
| 44 | OXC5671 | human | *Campylobacter jejuni* | 2 | 17 | 2 | 3 | 2 | 1 | 5 | 883 | ST-21 complex | 5 |
| 45 | OXC5691 | human | *Campylobacter jejuni* | 2 | 1 | 12 | 3 | 2 | 1 | 5 | 50 | ST-21 complex | 5 |
| 46 | OXC5696 | human | *Campylobacter jejuni* | 2 | 1 | 21 | 3 | 2 | 1 | 5 | 53 | ST-21 complex | 5 |
| 47 | OXC5708 | human | *Campylobacter jejuni* | 2 | 1 | 42 | 3 | 148 | 1 | 5 | 861 | ST-21 complex | 5 |
| 48 | OXC5710 | human | *Campylobacter jejuni* | 2 | 1 | 1 | 3 | 2 | 1 | 5 | 21 | ST-21 complex | 5 |
| 49 | OXC5713 | human | *Campylobacter jejuni* | 2 | 1 | 1 | 3 | 2 | 1 | 5 | 21 | ST-21 complex | 5 |
| 50 | OXC5720 | human | *Campylobacter jejuni* | 2 | 1 | 12 | 3 | 2 | 1 | 5 | 50 | ST-21 complex | 5 |
| 51 | OXC5495 | human | *Campylobacter jejuni* | 2 | 1 | 1 | 3 | 2 | 1 | 5 | 21 | ST-21 complex | 5 |
| 52 | OXC5617 | human | *Campylobacter jejuni* | 2 | 1 | 21 | 3 | 2 | 1 | 5 | 53 | ST-21 complex | 5 |
| 53 | OXC5618 | human | *Campylobacter jejuni* | 2 | 1 | 1 | 3 | 2 | 1 | 5 | 21 | ST-21 complex | 5 |
| 54 | OXC5625 | human | *Campylobacter jejuni* | 2 | 1 | 1 | 3 | 2 | 1 | 5 | 21 | ST-21 complex | 5 |
| 55 | OXC5628 | human | *Campylobacter jejuni* | 2 | 1 | 12 | 3 | 2 | 1 | 5 | 50 | ST-21 complex | 5 |
| 56 | OXC5639 | human | *Campylobacter jejuni* | 2 | 17 | 2 | 3 | 2 | 1 | 5 | 883 | ST-21 complex | 5 |
| 57 | OXC5647 | human | *Campylobacter jejuni* | 2 | 1 | 5 | 3 | 2 | 1 | 5 | 19 | ST-21 complex | 5 |
| 58 | OXC5649 | human | *Campylobacter jejuni* | 2 | 1 | 1 | 3 | 2 | 1 | 5 | 21 | ST-21 complex | 5 |

| | | | | | | | | | | | | |
|---|---|---|---|---|---|---|---|---|---|---|---|---|
| 59 | OXC5651 | human | *Campylobacter jejuni* | 2 | 1 | 12 | 3 | 2 | 1 | 5 | 50 | ST-21 complex | 5 |
| 60 | OXC5652 | human | *Campylobacter jejuni* | 2 | 1 | 1 | 3 | 2 | 1 | 5 | 21 | ST-21 complex | 5 |
| 61 | OXC5835 | human | *Campylobacter jejuni* | 2 | 1 | 1 | 3 | 2 | 1 | 5 | 21 | ST-21 complex | 5 |
| 62 | OXC5498 | human | *Campylobacter jejuni* | 2 | 1 | 1 | 3 | 2 | 1 | 5 | 21 | ST-21 complex | 5 |
| 63 | OXC5840 | human | *Campylobacter jejuni* | 2 | 1 | 5 | 3 | 2 | 1 | 5 | 19 | ST-21 complex | 5 |
| 64 | OXC5841 | human | *Campylobacter jejuni* | 8 | 1 | 6 | 3 | 2 | 1 | 12 | 2135 | ST-21 complex | 5 |
| 65 | OXC5843 | human | *Campylobacter jejuni* | 2 | 1 | 42 | 3 | 148 | 1 | 5 | 861 | ST-21 complex | 5 |
| 66 | OXC5846 | human | *Campylobacter jejuni* | 2 | 1 | 1 | 3 | 2 | 1 | 5 | 21 | ST-21 complex | 5 |
| 67 | OXC5848 | human | *Campylobacter jejuni* | 2 | 1 | 1 | 3 | 2 | 1 | 5 | 21 | ST-21 complex | 5 |
| 68 | OXC5862 | human | *Campylobacter jejuni* | 2 | 1 | 1 | 3 | 2 | 1 | 5 | 21 | ST-21 complex | 5 |
| 69 | OXC5871 | human | *Campylobacter jejuni* | 2 | 1 | 5 | 3 | 2 | 1 | 5 | 19 | ST-21 complex | 5 |
| 70 | OXC5879 | human | *Campylobacter jejuni* | 2 | 1 | 12 | 3 | 2 | 1 | 5 | 50 | ST-21 complex | 5 |
| 71 | OXC5880 | human | *Campylobacter jejuni* | 2 | 1 | 21 | 3 | 2 | 1 | 5 | 53 | ST-21 complex | 5 |
| 72 | OXC5413 | human | *Campylobacter jejuni* | 2 | 1 | 1 | 3 | 2 | 1 | 5 | 21 | ST-21 complex | 5 |
| 73 | OXC5414 | human | *Campylobacter jejuni* | 2 | 1 | 12 | 3 | 2 | 1 | 5 | 50 | ST-21 complex | 5 |
| 74 | OXC5416 | human | *Campylobacter jejuni* | 2 | 1 | 12 | 3 | 2 | 1 | 23 | 3574 | ST-21 complex | 5 |
| 75 | OXC5431 | human | *Campylobacter jejuni* | 2 | 1 | 1 | 3 | 492 | 1 | 5 | 5018 | ST-21 complex | 5 |
| 76 | OXC5435 | human | *Campylobacter jejuni* | 2 | 1 | 12 | 3 | 2 | 1 | 5 | 50 | ST-21 complex | 5 |
| 77 | OXC5437 | human | *Campylobacter jejuni* | 2 | 17 | 2 | 3 | 2 | 1 | 5 | 883 | ST-21 complex | 5 |
| 78 | OXC5438 | human | *Campylobacter jejuni* | 2 | 1 | 21 | 3 | 2 | 1 | 5 | 53 | ST-21 complex | 5 |
| 79 | OXC5444 | human | *Campylobacter jejuni* | 2 | 1 | 5 | 3 | 2 | 1 | 5 | 19 | ST-21 complex | 5 |
| 80 | OXC5445 | human | *Campylobacter jejuni* | 2 | 1 | 12 | 3 | 2 | 1 | 5 | 50 | ST-21 complex | 5 |
| 81 | OXC5451 | human | *Campylobacter jejuni* | 2 | 1 | 12 | 3 | 2 | 1 | 5 | 50 | ST-21 complex | 5 |
| 82 | OXC5462 | human | *Campylobacter jejuni* | 2 | 1 | 12 | 3 | 2 | 1 | 5 | 50 | ST-21 complex | 5 |
| 83 | OXC5463 | human | *Campylobacter jejuni* | 2 | 1 | 5 | 3 | 2 | 1 | 5 | 19 | ST-21 complex | 5 |
| 84 | OXC5465 | human | *Campylobacter jejuni* | 2 | 1 | 1 | 3 | 2 | 1 | 5 | 21 | ST-21 complex | 5 |
| 85 | OXC5470 | human | *Campylobacter jejuni* | 8 | 1 | 6 | 3 | 2 | 1 | 1 | 44 | ST-21 complex | 5 |
| 86 | OXC5724 | human | *Campylobacter jejuni* | 2 | 1 | 5 | 3 | 2 | 1 | 5 | 19 | ST-21 complex | 5 |
| 87 | OXC5725 | human | *Campylobacter jejuni* | 2 | 1 | 12 | 3 | 2 | 1 | 5 | 50 | ST-21 complex | 5 |
| 88 | OXC5728 | human | *Campylobacter jejuni* | 2 | 1 | 5 | 3 | 2 | 61 | 5 | 2355 | ST-21 complex | 5 |

| | | | | | | | | | | | | |
|---|---|---|---|---|---|---|---|---|---|---|---|---|
| 89 | OXC5731 | human | *Campylobacter jejuni* | 2 | 1 | 12 | 3 | 2 | 1 | 5 | 50 | ST-21 complex | 5 |
| 90 | OXC5737 | human | *Campylobacter jejuni* | 2 | 1 | 5 | 3 | 2 | 54 | 5 | 266 | ST-21 complex | 5 |
| 91 | OXC5739 | human | *Campylobacter jejuni* | 2 | 1 | 1 | 3 | 2 | 1 | 5 | 21 | ST-21 complex | 5 |
| 92 | OXC5754 | human | *Campylobacter jejuni* | 2 | 1 | 1 | 3 | 2 | 1 | 5 | 21 | ST-21 complex | 5 |
| 93 | OXC5757 | human | *Campylobacter jejuni* | 2 | 1 | 1 | 3 | 2 | 1 | 5 | 21 | ST-21 complex | 5 |
| 94 | OXC5766 | human | *Campylobacter jejuni* | 2 | 1 | 12 | 3 | 2 | 1 | 5 | 50 | ST-21 complex | 5 |
| 95 | OXC5767 | human | *Campylobacter jejuni* | 2 | 1 | 21 | 3 | 2 | 1 | 5 | 53 | ST-21 complex | 5 |
| 96 | OXC5771 | human | *Campylobacter jejuni* | 2 | 1 | 12 | 3 | 2 | 1 | 5 | 50 | ST-21 complex | 5 |
| 97 | OXC5772 | human | *Campylobacter jejuni* | 8 | 1 | 6 | 3 | 2 | 1 | 12 | 2135 | ST-21 complex | 5 |
| 98 | OXC5774 | human | *Campylobacter jejuni* | 2 | 1 | 5 | 3 | 2 | 3 | 5 | 190 | ST-21 complex | 5 |
| 99 | OXC5791 | human | *Campylobacter jejuni* | 2 | 1 | 1 | 3 | 2 | 1 | 5 | 21 | ST-21 complex | 5 |
| 100 | OXC5799 | human | *Campylobacter jejuni* | 2 | 1 | 12 | 3 | 2 | 1 | 5 | 50 | ST-21 complex | 5 |
| 101 | OXC5804 | human | *Campylobacter jejuni* | 2 | 1 | 21 | 3 | 2 | 1 | 5 | 53 | ST-21 complex | 5 |
| 102 | OXC5808 | human | *Campylobacter jejuni* | 2 | 1 | 21 | 3 | 2 | 1 | 5 | 53 | ST-21 complex | 5 |
| 103 | OXC5811 | human | *Campylobacter jejuni* | 8 | 1 | 6 | 3 | 2 | 1 | 12 | 2135 | ST-21 complex | 5 |
| 104 | OXC5818 | human | *Campylobacter jejuni* | 2 | 1 | 5 | 3 | 2 | 1 | 5 | 19 | ST-21 complex | 5 |
| 105 | OXC5819 | human | *Campylobacter jejuni* | 2 | 1 | 1 | 3 | 2 | 1 | 5 | 21 | ST-21 complex | 5 |
| 106 | OXC5821 | human | *Campylobacter jejuni* | 8 | 1 | 6 | 3 | 2 | 1 | 12 | 2135 | ST-21 complex | 5 |
| 107 | OXC5822 | human | *Campylobacter jejuni* | 2 | 1 | 1 | 3 | 2 | 1 | 5 | 21 | ST-21 complex | 5 |
| 108 | OXC5823 | human | *Campylobacter jejuni* | 2 | 1 | 21 | 3 | 2 | 1 | 5 | 53 | ST-21 complex | 5 |
| 109 | OXC5825 | human | *Campylobacter jejuni* | 2 | 1 | 21 | 3 | 2 | 1 | 5 | 53 | ST-21 complex | 5 |
| 110 | OXC5826 | human | *Campylobacter jejuni* | 2 | 1 | 1 | 3 | 2 | 1 | 5 | 21 | ST-21 complex | 5 |
| 111 | OXC5829 | human | *Campylobacter jejuni* | 2 | 1 | 12 | 3 | 2 | 1 | 5 | 50 | ST-21 complex | 5 |
| 112 | OXC5832 | human | *Campylobacter jejuni* | 2 | 1 | 1 | 3 | 2 | 1 | 5 | 21 | ST-21 complex | 5 |
| 113 | OXC5887 | human | *Campylobacter jejuni* | 2 | 1 | 5 | 3 | 2 | 1 | 5 | 19 | ST-21 complex | 5 |
| 114 | OXC5888 | human | *Campylobacter jejuni* | 8 | 1 | 6 | 3 | 2 | 1 | 12 | 2135 | ST-21 complex | 5 |
| 115 | OXC5895 | human | *Campylobacter jejuni* | 8 | 1 | 6 | 3 | 2 | 1 | 12 | 2135 | ST-21 complex | 5 |
| 116 | OXC5899 | human | *Campylobacter jejuni* | 2 | 1 | 5 | 3 | 2 | 1 | 5 | 19 | ST-21 complex | 5 |
| 117 | OXC5902 | human | *Campylobacter jejuni* | 2 | 1 | 12 | 3 | 2 | 1 | 5 | 50 | ST-21 complex | 5 |
| 118 | OXC5905 | human | *Campylobacter jejuni* | 2 | 1 | 12 | 3 | 2 | 1 | 5 | 50 | ST-21 complex | 5 |

| | | | | | | | | | | | | |
|---|---|---|---|---|---|---|---|---|---|---|---|---|
| 119 | OXC5906 | human | *Campylobacter jejuni* | 2 | 1 | 12 | 3 | 2 | 1 | 5 | 50 | ST-21 complex | 5 |
| 120 | OXC5909 | human | *Campylobacter jejuni* | 2 | 1 | 21 | 3 | 2 | 1 | 5 | 53 | ST-21 complex | 5 |
| 121 | OXC5918 | human | *Campylobacter jejuni* | 2 | 1 | 21 | 3 | 2 | 1 | 5 | 53 | ST-21 complex | 5 |
| 122 | OXC5937 | human | *Campylobacter jejuni* | 2 | 1 | 1 | 3 | 2 | 1 | 5 | 21 | ST-21 complex | 5 |
| 123 | OXC5924 | human | *Campylobacter jejuni* | 2 | 1 | 1 | 3 | 2 | 1 | 5 | 21 | ST-21 complex | 5 |
| 124 | OXC5925 | human | *Campylobacter jejuni* | 2 | 1 | 12 | 3 | 2 | 1 | 5 | 50 | ST-21 complex | 5 |
| 125 | OXC5926 | human | *Campylobacter jejuni* | 2 | 1 | 12 | 3 | 2 | 1 | 5 | 50 | ST-21 complex | 5 |
| 126 | OXC5929 | human | *Campylobacter jejuni* | 2 | 1 | 1 | 3 | 2 | 1 | 5 | 21 | ST-21 complex | 5 |
| 127 | OXC6548 | human | *Campylobacter jejuni* | 2 | 1 | 5 | 3 | 2 | 1 | 5 | 19 | ST-21 complex | 5 |
| 128 | OXC6552 | human | *Campylobacter jejuni* | 2 | 1 | 12 | 3 | 2 | 1 | 12 | 3769 | ST-21 complex | 5 |
| 129 | OXC6558 | human | *Campylobacter jejuni* | 2 | 1 | 21 | 3 | 2 | 1 | 5 | 53 | ST-21 complex | 5 |
| 130 | OXC6562 | human | *Campylobacter jejuni* | 2 | 1 | 1 | 3 | 2 | 1 | 5 | 21 | ST-21 complex | 5 |
| 131 | OXC6563 | human | *Campylobacter jejuni* | 2 | 1 | 1 | 3 | 2 | 1 | 5 | 21 | ST-21 complex | 5 |
| 132 | OXC6564 | human | *Campylobacter jejuni* | 2 | 1 | 1 | 3 | 2 | 1 | 5 | 21 | ST-21 complex | 5 |
| 133 | OXC6565 | human | *Campylobacter jejuni* | 2 | 1 | 12 | 3 | 2 | 1 | 5 | 50 | ST-21 complex | 5 |
| 134 | OXC6571 | human | *Campylobacter jejuni* | 2 | 1 | 12 | 3 | 2 | 1 | 5 | 50 | ST-21 complex | 5 |
| 135 | OXC6573 | human | *Campylobacter jejuni* | 2 | 1 | 1 | 3 | 2 | 1 | 5 | 21 | ST-21 complex | 5 |
| 136 | OXC6251 | human | *Campylobacter jejuni* | 2 | 1 | 1 | 3 | 2 | 1 | 5 | 21 | ST-21 complex | 5 |
| 137 | OXC6257 | human | *Campylobacter jejuni* | 2 | 1 | 1 | 3 | 2 | 1 | 5 | 21 | ST-21 complex | 5 |
| 138 | OXC6266 | human | *Campylobacter jejuni* | 2 | 1 | 12 | 3 | 2 | 1 | 5 | 50 | ST-21 complex | 5 |
| 139 | OXC6270 | human | *Campylobacter jejuni* | 2 | 1 | 1 | 3 | 2 | 1 | 5 | 21 | ST-21 complex | 5 |
| 140 | OXC6275 | human | *Campylobacter jejuni* | 2 | 1 | 1 | 3 | 2 | 1 | 5 | 21 | ST-21 complex | 5 |
| 141 | OXC6277 | human | *Campylobacter jejuni* | 2 | 1 | 12 | 3 | 2 | 1 | 5 | 50 | ST-21 complex | 5 |
| 142 | OXC6282 | human | *Campylobacter jejuni* | 2 | 1 | 21 | 3 | 2 | 1 | 5 | 53 | ST-21 complex | 5 |
| 143 | OXC6285 | human | *Campylobacter jejuni* | 2 | 1 | 1 | 3 | 2 | 1 | 5 | 21 | ST-21 complex | 5 |
| 144 | OXC6286 | human | *Campylobacter jejuni* | 2 | 1 | 12 | 3 | 2 | 1 | 5 | 50 | ST-21 complex | 5 |
| 145 | OXC6289 | human | *Campylobacter jejuni* | 2 | 1 | 12 | 3 | 2 | 1 | 5 | 50 | ST-21 complex | 5 |
| 146 | OXC6292 | human | *Campylobacter jejuni* | 2 | 1 | 12 | 3 | 2 | 1 | 5 | 50 | ST-21 complex | 5 |
| 147 | OXC6300 | human | *Campylobacter jejuni* | 8 | 1 | 6 | 3 | 2 | 1 | 12 | 2135 | ST-21 complex | 5 |
| 148 | OXC6301 | human | *Campylobacter jejuni* | 2 | 1 | 42 | 3 | 148 | 1 | 5 | 861 | ST-21 complex | 5 |

| | | | | | | | | | | | | |
|---|---|---|---|---|---|---|---|---|---|---|---|---|
| 149 | OXC6303 | human | *Campylobacter jejuni* | 2 | 1 | 5 | 3 | 2 | 1 | 5 | 19 | ST-21 complex | 5 |
| 150 | OXC6305 | human | *Campylobacter jejuni* | 2 | 1 | 42 | 3 | 148 | 1 | 5 | 861 | ST-21 complex | 5 |
| 151 | OXC6310 | human | *Campylobacter jejuni* | 2 | 1 | 1 | 3 | 2 | 1 | 5 | 21 | ST-21 complex | 5 |
| 152 | OXC6317 | human | *Campylobacter jejuni* | 2 | 1 | 1 | 3 | 2 | 1 | 5 | 21 | ST-21 complex | 5 |
| 153 | OXC6324 | human | *Campylobacter jejuni* | 2 | 1 | 21 | 3 | 2 | 1 | 5 | 53 | ST-21 complex | 5 |
| 154 | OXC6325 | human | *Campylobacter jejuni* | 2 | 1 | 5 | 3 | 2 | 1 | 5 | 19 | ST-21 complex | 5 |
| 155 | OXC6326 | human | *Campylobacter jejuni* | 2 | 1 | 12 | 3 | 2 | 1 | 5 | 50 | ST-21 complex | 5 |
| 156 | OXC6329 | human | *Campylobacter jejuni* | 2 | 1 | 1 | 3 | 2 | 1 | 5 | 21 | ST-21 complex | 5 |
| 157 | OXC6331 | human | *Campylobacter jejuni* | 2 | 1 | 12 | 3 | 2 | 1 | 5 | 50 | ST-21 complex | 5 |
| 158 | OXC6334 | human | *Campylobacter jejuni* | 2 | 1 | 1 | 3 | 2 | 1 | 5 | 21 | ST-21 complex | 5 |
| 159 | OXC6335 | human | *Campylobacter jejuni* | 2 | 1 | 1 | 3 | 2 | 1 | 5 | 21 | ST-21 complex | 5 |
| 160 | OXC6340 | human | *Campylobacter jejuni* | 2 | 1 | 1 | 3 | 2 | 1 | 3 | 262 | ST-21 complex | 5 |
| 161 | OXC6347 | human | *Campylobacter jejuni* | 2 | 1 | 12 | 3 | 2 | 1 | 5 | 50 | ST-21 complex | 5 |
| 162 | OXC6355 | human | *Campylobacter jejuni* | 2 | 1 | 5 | 3 | 2 | 1 | 5 | 19 | ST-21 complex | 5 |
| 163 | OXC6367 | human | *Campylobacter jejuni* | 2 | 1 | 21 | 3 | 2 | 1 | 5 | 53 | ST-21 complex | 5 |
| 164 | OXC6370 | human | *Campylobacter jejuni* | 2 | 1 | 5 | 3 | 2 | 1 | 5 | 19 | ST-21 complex | 5 |
| 165 | OXC6379 | human | *Campylobacter jejuni* | 2 | 1 | 5 | 462 | 2 | 1 | 5 | 5726 | ST-21 complex | 5 |
| 166 | OXC6383 | human | *Campylobacter jejuni* | 2 | 1 | 1 | 3 | 2 | 1 | 5 | 21 | ST-21 complex | 5 |
| 167 | OXC6384 | human | *Campylobacter jejuni* | 2 | 1 | 5 | 3 | 2 | 1 | 5 | 19 | ST-21 complex | 5 |
| 168 | OXC6393 | human | *Campylobacter jejuni* | 2 | 1 | 12 | 462 | 2 | 1 | 5 | 5727 | ST-21 complex | 5 |
| 169 | OXC6394 | human | *Campylobacter jejuni* | 2 | 1 | 5 | 3 | 2 | 1 | 5 | 19 | ST-21 complex | 5 |
| 170 | OXC6405 | human | *Campylobacter jejuni* | 2 | 1 | 1 | 3 | 2 | 1 | 5 | 21 | ST-21 complex | 5 |
| 171 | OXC6420 | human | *Campylobacter jejuni* | 2 | 1 | 1 | 3 | 2 | 1 | 5 | 21 | ST-21 complex | 5 |
| 172 | OXC6429 | human | *Campylobacter jejuni* | 2 | 1 | 1 | 3 | 492 | 1 | 5 | 5018 | ST-21 complex | 5 |
| 173 | OXC6449 | human | *Campylobacter jejuni* | 2 | 1 | 12 | 3 | 2 | 1 | 5 | 50 | ST-21 complex | 5 |
| 174 | OXC6457 | human | *Campylobacter jejuni* | 2 | 1 | 12 | 3 | 2 | 1 | 5 | 50 | ST-21 complex | 5 |
| 175 | OXC6459 | human | *Campylobacter jejuni* | 2 | 1 | 12 | 3 | 2 | 1 | 5 | 50 | ST-21 complex | 5 |
| 176 | OXC6461 | human | *Campylobacter jejuni* | 2 | 1 | 12 | 3 | 2 | 1 | 5 | 50 | ST-21 complex | 5 |
| 177 | OXC6464 | human | *Campylobacter jejuni* | 2 | 1 | 1 | 3 | 492 | 1 | 5 | 5018 | ST-21 complex | 5 |
| 178 | OXC6479 | human | *Campylobacter jejuni* | 2 | 17 | 2 | 3 | 2 | 1 | 5 | 883 | ST-21 complex | 5 |

| | | | | | | | | | | | | |
|---|---|---|---|---|---|---|---|---|---|---|---|---|
| 179 | OXC6483 | human | *Campylobacter jejuni* | 2 | 1 | 1 | 3 | 2 | 1 | 5 | 21 | ST-21 complex | 5 |
| 180 | OXC6489 | human | *Campylobacter jejuni* | 2 | 1 | 12 | 3 | 2 | 1 | 5 | 50 | ST-21 complex | 5 |
| 181 | OXC6493 | human | *Campylobacter jejuni* | 2 | 1 | 12 | 3 | 2 | 1 | 5 | 50 | ST-21 complex | 5 |
| 182 | OXC6496 | human | *Campylobacter jejuni* | 2 | 1 | 1 | 3 | 2 | 1 | 5 | 21 | ST-21 complex | 5 |
| 183 | OXC6500 | human | *Campylobacter jejuni* | 8 | 1 | 6 | 3 | 2 | 1 | 12 | 2135 | ST-21 complex | 5 |
| 184 | OXC6502 | human | *Campylobacter jejuni* | 2 | 1 | 12 | 3 | 2 | 1 | 5 | 50 | ST-21 complex | 5 |
| 185 | OXC6505 | human | *Campylobacter jejuni* | 2 | 1 | 10 | 3 | 2 | 1 | 5 | 141 | ST-21 complex | 5 |
| 186 | OXC6508 | human | *Campylobacter jejuni* | 2 | 1 | 1 | 3 | 2 | 1 | 5 | 21 | ST-21 complex | 5 |
| 187 | OXC6514 | human | *Campylobacter jejuni* | 2 | 84 | 12 | 3 | 11 | 1 | 5 | 3102 | ST-21 complex | 5 |
| 188 | OXC6516 | human | *Campylobacter jejuni* | 2 | 1 | 21 | 3 | 2 | 1 | 5 | 53 | ST-21 complex | 5 |
| 189 | OXC6519 | human | *Campylobacter jejuni* | 2 | 1 | 1 | 3 | 2 | 1 | 5 | 21 | ST-21 complex | 5 |
| 190 | OXC6521 | human | *Campylobacter jejuni* | 2 | 1 | 1 | 5 | 2 | 1 | 5 | 47 | ST-21 complex | 5 |
| 191 | OXC6524 | human | *Campylobacter jejuni* | 2 | 1 | 12 | 3 | 2 | 1 | 5 | 50 | ST-21 complex | 5 |
| 192 | OXC6527 | human | *Campylobacter jejuni* | 2 | 1 | 12 | 3 | 2 | 1 | 5 | 50 | ST-21 complex | 5 |
| 193 | OXC6530 | human | *Campylobacter jejuni* | 2 | 1 | 12 | 3 | 2 | 1 | 5 | 50 | ST-21 complex | 5 |
| 194 | OXC6531 | human | *Campylobacter jejuni* | 2 | 1 | 12 | 3 | 2 | 1 | 5 | 50 | ST-21 complex | 5 |
| 195 | OXC6538 | human | *Campylobacter jejuni* | 2 | 1 | 5 | 3 | 2 | 1 | 5 | 19 | ST-21 complex | 5 |
| 196 | OXC6539 | human | *Campylobacter jejuni* | 2 | 1 | 1 | 3 | 2 | 1 | 5 | 21 | ST-21 complex | 5 |
| 197 | OXC6543 | human | *Campylobacter jejuni* | 2 | 1 | 12 | 3 | 2 | 1 | 5 | 50 | ST-21 complex | 5 |
| 198 | OXC6255 | human | *Campylobacter jejuni* | 2 | 1 | 1 | 3 | 2 | 1 | 5 | 21 | ST-21 complex | 5 |
| 199 | OXC6590 | human | *Campylobacter jejuni* | 2 | 1 | 12 | 3 | 2 | 1 | 5 | 50 | ST-21 complex | 5 |
| 200 | OXC6596 | human | *Campylobacter jejuni* | 2 | 1 | 1 | 3 | 2 | 1 | 5 | 21 | ST-21 complex | 5 |
| 201 | OXC6598 | human | *Campylobacter jejuni* | 2 | 1 | 12 | 3 | 2 | 1 | 5 | 50 | ST-21 complex | 5 |
| 202 | OXC6600 | human | *Campylobacter jejuni* | 2 | 1 | 12 | 3 | 2 | 1 | 5 | 50 | ST-21 complex | 5 |
| 203 | OXC6602 | human | *Campylobacter jejuni* | 2 | 1 | 1 | 3 | 2 | 1 | 5 | 21 | ST-21 complex | 5 |
| 204 | OXC6603 | human | *Campylobacter jejuni* | 2 | 1 | 5 | 3 | 2 | 1 | 5 | 19 | ST-21 complex | 5 |
| 205 | OXC6604 | human | *Campylobacter jejuni* | 2 | 1 | 1 | 3 | 2 | 1 | 5 | 21 | ST-21 complex | 5 |
| 206 | OXC6613 | human | *Campylobacter jejuni* | 2 | 1 | 12 | 3 | 2 | 1 | 5 | 50 | ST-21 complex | 5 |
| 207 | OXC6615 | human | *Campylobacter jejuni* | 2 | 1 | 12 | 3 | 2 | 1 | 5 | 50 | ST-21 complex | 5 |
| 208 | OXC6616 | human | *Campylobacter jejuni* | 2 | 1 | 1 | 3 | 2 | 1 | 5 | 21 | ST-21 complex | 5 |

| | | | | | | | | | | | | |
|---|---|---|---|---|---|---|---|---|---|---|---|---|
| 209 | OXC6619 | human | *Campylobacter jejuni* | 2 | 1 | 1 | 3 | 2 | 1 | 5 | 21 | ST-21 complex | 5 |
| 210 | OXC6625 | human | *Campylobacter jejuni* | 2 | 1 | 1 | 3 | 2 | 1 | 5 | 21 | ST-21 complex | 5 |
| 211 | OXC6627 | human | *Campylobacter jejuni* | 2 | 1 | 1 | 3 | 2 | 1 | 5 | 21 | ST-21 complex | 5 |
| 212 | OXC6629 | human | *Campylobacter jejuni* | 2 | 1 | 1 | 3 | 2 | 1 | 5 | 21 | ST-21 complex | 5 |
| 213 | OXC6633 | human | *Campylobacter jejuni* | 2 | 1 | 21 | 3 | 2 | 1 | 5 | 53 | ST-21 complex | 5 |
| 214 | OXC6636 | human | *Campylobacter jejuni* | 2 | 1 | 12 | 3 | 2 | 1 | 5 | 50 | ST-21 complex | 5 |
| 215 | OXC6637 | human | *Campylobacter jejuni* | 2 | 1 | 1 | 3 | 2 | 1 | 5 | 21 | ST-21 complex | 5 |
| 216 | OXC6642 | human | *Campylobacter jejuni* | 2 | 1 | 1 | 3 | 2 | 1 | 5 | 21 | ST-21 complex | 5 |
| 217 | OXC6643 | human | *Campylobacter jejuni* | 2 | 1 | 21 | 3 | 2 | 1 | 5 | 53 | ST-21 complex | 5 |
| 218 | OXC6656 | human | *Campylobacter jejuni* | 2 | 1 | 21 | 3 | 2 | 1 | 5 | 53 | ST-21 complex | 5 |
| 219 | OXC6664 | human | *Campylobacter jejuni* | 2 | 1 | 12 | 3 | 2 | 1 | 5 | 50 | ST-21 complex | 5 |
| 220 | OXC6665 | human | *Campylobacter jejuni* | 2 | 1 | 12 | 3 | 2 | 1 | 5 | 50 | ST-21 complex | 5 |
| 221 | OXC6666 | human | *Campylobacter jejuni* | 2 | 1 | 12 | 3 | 2 | 1 | 5 | 50 | ST-21 complex | 5 |
| 222 | OXC6667 | human | *Campylobacter jejuni* | 8 | 1 | 6 | 3 | 2 | 1 | 1 | 44 | ST-21 complex | 5 |
| 223 | OXC6671 | human | *Campylobacter jejuni* | 2 | 1 | 1 | 3 | 2 | 1 | 5 | 21 | ST-21 complex | 5 |
| 224 | OXC6672 | human | *Campylobacter jejuni* | 2 | 1 | 1 | 5 | 2 | 1 | 5 | 47 | ST-21 complex | 5 |
| 225 | OXC6673 | human | *Campylobacter jejuni* | 2 | 1 | 12 | 3 | 2 | 1 | 5 | 50 | ST-21 complex | 5 |
| 226 | OXC6677 | human | *Campylobacter jejuni* | 2 | 1 | 12 | 3 | 2 | 1 | 5 | 50 | ST-21 complex | 5 |
| 227 | OXC6678 | human | *Campylobacter jejuni* | 2 | 1 | 1 | 3 | 2 | 1 | 5 | 21 | ST-21 complex | 5 |
| 228 | OXC6681 | human | *Campylobacter jejuni* | 2 | 1 | 1 | 3 | 2 | 1 | 5 | 21 | ST-21 complex | 5 |
| 229 | OXC6682 | human | *Campylobacter jejuni* | 2 | 1 | 21 | 3 | 2 | 1 | 5 | 53 | ST-21 complex | 5 |
| 230 | OXC6689 | human | *Campylobacter jejuni* | 2 | 1 | 1 | 3 | 2 | 1 | 5 | 21 | ST-21 complex | 5 |
| 231 | OXC6690 | human | *Campylobacter jejuni* | 2 | 1 | 1 | 3 | 2 | 1 | 5 | 21 | ST-21 complex | 5 |
| 232 | OXC6691 | human | *Campylobacter jejuni* | 2 | 1 | 5 | 3 | 2 | 3 | 5 | 190 | ST-21 complex | 5 |
| 233 | OXC6702 | human | *Campylobacter jejuni* | 2 | 1 | 1 | 3 | 2 | 1 | 5 | 21 | ST-21 complex | 5 |
| 234 | OXC6703 | human | *Campylobacter jejuni* | 2 | 1 | 5 | 3 | 2 | 1 | 5 | 19 | ST-21 complex | 5 |
| 235 | OXC6706 | human | *Campylobacter jejuni* | 2 | 1 | 5 | 3 | 2 | 1 | 5 | 19 | ST-21 complex | 5 |
| 236 | OXC6711 | human | *Campylobacter jejuni* | 2 | 1 | 1 | 3 | 2 | 1 | 5 | 21 | ST-21 complex | 5 |
| 237 | OXC6712 | human | *Campylobacter jejuni* | 2 | 1 | 1 | 3 | 2 | 1 | 5 | 21 | ST-21 complex | 5 |
| 238 | OXC6713 | human | *Campylobacter jejuni* | 2 | 1 | 12 | 3 | 2 | 1 | 5 | 50 | ST-21 complex | 5 |

| | | | | | | | | | | | | |
|---|---|---|---|---|---|---|---|---|---|---|---|---|
| 239 | OXC6714 | human | *Campylobacter jejuni* | 2 | 1 | 12 | 3 | 2 | 1 | 5 | 50 | ST-21 complex | 5 |
| 240 | OXC6727 | human | *Campylobacter jejuni* | 2 | 2 | 52 | 3 | 2 | 100 | 5 | 5811 | ST-21 complex | 5 |
| 241 | OXC6728 | human | *Campylobacter jejuni* | 2 | 1 | 1 | 3 | 2 | 1 | 5 | 21 | ST-21 complex | 5 |
| 242 | OXC6730 | human | *Campylobacter jejuni* | 2 | 1 | 1 | 3 | 2 | 1 | 5 | 21 | ST-21 complex | 5 |
| 243 | OXC6731 | human | *Campylobacter jejuni* | 2 | 1 | 1 | 3 | 2 | 1 | 5 | 21 | ST-21 complex | 5 |
| 244 | OXC6734 | human | *Campylobacter jejuni* | 2 | 1 | 1 | 3 | 2 | 1 | 5 | 21 | ST-21 complex | 5 |
| 245 | OXC6741 | human | *Campylobacter jejuni* | 2 | 1 | 21 | 3 | 2 | 1 | 5 | 53 | ST-21 complex | 5 |
| 246 | OXC6745 | human | *Campylobacter jejuni* | 2 | 1 | 1 | 3 | 2 | 1 | 5 | 21 | ST-21 complex | 5 |
| 247 | OXC6751 | human | *Campylobacter jejuni* | 8 | 1 | 6 | 3 | 2 | 1 | 12 | 2135 | ST-21 complex | 5 |
| 248 | OXC6752 | human | *Campylobacter jejuni* | 8 | 1 | 6 | 3 | 2 | 1 | 12 | 2135 | ST-21 complex | 5 |
| 249 | OXC6763 | human | *Campylobacter jejuni* | 2 | 1 | 12 | 3 | 2 | 1 | 5 | 50 | ST-21 complex | 5 |
| 250 | OXC6764 | human | *Campylobacter jejuni* | 2 | 1 | 1 | 3 | 2 | 1 | 5 | 21 | ST-21 complex | 5 |
| 251 | OXC6766 | human | *Campylobacter jejuni* | 2 | 1 | 1 | 3 | 2 | 1 | 3 | 262 | ST-21 complex | 5 |
| 252 | OXC6779 | human | *Campylobacter jejuni* | 2 | 1 | 21 | 3 | 2 | 1 | 5 | 53 | ST-21 complex | 5 |
| 253 | OXC6781 | human | *Campylobacter jejuni* | 8 | 1 | 6 | 3 | 2 | 1 | 12 | 2135 | ST-21 complex | 5 |
| 254 | OXC6786 | human | *Campylobacter jejuni* | 8 | 1 | 6 | 3 | 2 | 1 | 12 | 2135 | ST-21 complex | 5 |
| 255 | OXC6789 | human | *Campylobacter jejuni* | 2 | 1 | 21 | 3 | 2 | 1 | 5 | 53 | ST-21 complex | 5 |
| 256 | OXC6791 | human | *Campylobacter jejuni* | 2 | 1 | 1 | 3 | 2 | 1 | 5 | 21 | ST-21 complex | 5 |
| 257 | OXC6798 | human | *Campylobacter jejuni* | 2 | 17 | 2 | 3 | 2 | 1 | 301 | 4526 | ST-21 complex | 5 |
| 258 | OXC6804 | human | *Campylobacter jejuni* | 2 | 1 | 1 | 3 | 2 | 1 | 5 | 21 | ST-21 complex | 5 |
| 259 | OXC6812 | human | *Campylobacter jejuni* | 2 | 1 | 1 | 3 | 2 | 1 | 5 | 21 | ST-21 complex | 5 |
| 260 | OXC6815 | human | *Campylobacter jejuni* | 2 | 1 | 1 | 3 | 2 | 1 | 5 | 21 | ST-21 complex | 5 |
| 261 | OXC6820 | human | *Campylobacter jejuni* | 2 | 1 | 12 | 3 | 2 | 1 | 5 | 50 | ST-21 complex | 5 |
| 262 | OXC6823 | human | *Campylobacter jejuni* | 2 | 17 | 2 | 3 | 2 | 1 | 5 | 883 | ST-21 complex | 5 |
| 263 | OXC6824 | human | *Campylobacter jejuni* | 2 | 1 | 1 | 3 | 2 | 1 | 3 | 262 | ST-21 complex | 5 |
| 264 | OXC6833 | human | *Campylobacter jejuni* | 2 | 1 | 1 | 3 | 7 | 1 | 5 | 104 | ST-21 complex | 5 |
| 265 | OXC6932 | human | *Campylobacter jejuni* | 2 | 1 | 12 | 3 | 2 | 1 | 5 | 50 | ST-21 complex | 5 |
| 266 | OXC6941 | human | *Campylobacter jejuni* | 2 | 1 | 21 | 3 | 2 | 1 | 5 | 53 | ST-21 complex | 5 |
| 267 | OXC6945 | human | *Campylobacter jejuni* | 2 | 1 | 1 | 3 | 2 | 1 | 5 | 21 | ST-21 complex | 5 |
| 268 | OXC6946 | human | *Campylobacter jejuni* | 2 | 1 | 12 | 3 | 2 | 1 | 5 | 50 | ST-21 complex | 5 |

| | | | | | | | | | | | | |
|---|---|---|---|---|---|---|---|---|---|---|---|---|
| 269 | OXC6948 | human | *Campylobacter jejuni* | 2 | 1 | 1 | 3 | 2 | 1 | 5 | 21 | ST-21 complex | 5 |
| 270 | OXC6949 | human | *Campylobacter jejuni* | 2 | 1 | 12 | 3 | 2 | 1 | 5 | 50 | ST-21 complex | 5 |
| 271 | OXC6953 | human | *Campylobacter jejuni* | 2 | 1 | 1 | 3 | 2 | 1 | 5 | 21 | ST-21 complex | 5 |
| 272 | OXC6956 | human | *Campylobacter jejuni* | 2 | 1 | 12 | 3 | 2 | 1 | 5 | 50 | ST-21 complex | 5 |
| 273 | OXC6959 | human | *Campylobacter jejuni* | 2 | 1 | 12 | 462 | 2 | 1 | 5 | 5727 | ST-21 complex | 5 |
| 274 | OXC6961 | human | *Campylobacter jejuni* | 2 | 1 | 1 | 3 | 2 | 1 | 5 | 21 | ST-21 complex | 5 |
| 275 | OXC6964 | human | *Campylobacter jejuni* | 2 | 1 | 12 | 3 | 2 | 1 | 5 | 50 | ST-21 complex | 5 |
| 276 | OXC6979 | human | *Campylobacter jejuni* | 2 | 1 | 1 | 3 | 2 | 1 | 5 | 21 | ST-21 complex | 5 |
| 277 | OXC6981 | human | *Campylobacter jejuni* | 8 | 1 | 6 | 3 | 2 | 1 | 1 | 44 | ST-21 complex | 5 |
| 278 | OXC6985 | human | *Campylobacter jejuni* | 2 | 17 | 12 | 462 | 2 | 1 | 5 | 6135 | ST-21 complex | 5 |
| 279 | OXC6988 | human | *Campylobacter jejuni* | 2 | 1 | 5 | 462 | 2 | 1 | 5 | 5726 | ST-21 complex | 5 |
| 280 | OXC6994 | human | *Campylobacter jejuni* | 2 | 1 | 12 | 3 | 2 | 1 | 5 | 50 | ST-21 complex | 5 |
| 281 | OXC6998 | human | *Campylobacter jejuni* | 2 | 1 | 1 | 3 | 2 | 1 | 5 | 21 | ST-21 complex | 5 |
| 282 | OXC7000 | human | *Campylobacter jejuni* | 2 | 1 | 1 | 3 | 2 | 1 | 5 | 21 | ST-21 complex | 5 |
| 283 | OXC7104 | human | *Campylobacter jejuni* | 2 | 1 | 5 | 10 | 2 | 1 | 6 | 6137 | ST-21 complex | 5 |
| 284 | OXC7105 | human | *Campylobacter jejuni* | 2 | 1 | 5 | 10 | 2 | 1 | 6 | 6137 | ST-21 complex | 5 |
| 285 | OXC7118 | human | *Campylobacter jejuni* | 2 | 1 | 5 | 3 | 2 | 54 | 5 | 266 | ST-21 complex | 5 |
| 286 | OXC7132 | human | *Campylobacter jejuni* | 2 | 1 | 12 | 3 | 2 | 1 | 5 | 50 | ST-21 complex | 5 |
| 287 | OXC7134 | human | *Campylobacter jejuni* | 2 | 1 | 1 | 3 | 2 | 1 | 5 | 21 | ST-21 complex | 5 |
| 288 | OXC7136 | human | *Campylobacter jejuni* | 2 | 1 | 1 | 3 | 2 | 1 | 5 | 21 | ST-21 complex | 5 |
| 289 | OXC7137 | human | *Campylobacter jejuni* | 2 | 1 | 1 | 3 | 2 | 1 | 5 | 21 | ST-21 complex | 5 |
| 290 | OXC7140 | human | *Campylobacter jejuni* | 2 | 1 | 1 | 3 | 2 | 1 | 5 | 21 | ST-21 complex | 5 |
| 291 | OXC7141 | human | *Campylobacter jejuni* | 8 | 1 | 6 | 3 | 2 | 1 | 1 | 44 | ST-21 complex | 5 |
| 292 | OXC7142 | human | *Campylobacter jejuni* | 2 | 1 | 1 | 3 | 2 | 1 | 5 | 21 | ST-21 complex | 5 |
| 293 | OXC7147 | human | *Campylobacter jejuni* | 2 | 1 | 1 | 3 | 2 | 1 | 5 | 21 | ST-21 complex | 5 |
| 294 | OXC7148 | human | *Campylobacter jejuni* | 2 | 1 | 79 | 3 | 2 | 1 | 5 | 822 | ST-21 complex | 5 |
| 295 | OXC7150 | human | *Campylobacter jejuni* | 8 | 1 | 6 | 3 | 2 | 1 | 1 | 44 | ST-21 complex | 5 |
| 296 | OXC7153 | human | *Campylobacter jejuni* | 2 | 1 | 1 | 3 | 2 | 1 | 5 | 21 | ST-21 complex | 5 |
| 297 | OXC7160 | human | *Campylobacter jejuni* | 2 | 1 | 5 | 3 | 2 | 1 | 5 | 19 | ST-21 complex | 5 |
| 298 | OXC7163 | human | *Campylobacter jejuni* | 2 | 1 | 5 | 3 | 2 | 1 | 5 | 19 | ST-21 complex | 5 |

| | | | | | | | | | | | | |
|---|---|---|---|---|---|---|---|---|---|---|---|---|
| 299 | OXC7168 | human | *Campylobacter jejuni* | 2 | 1 | 5 | 3 | 2 | 1 | 5 | 19 | ST-21 complex | 5 |
| 300 | OXC6836 | human | *Campylobacter jejuni* | 2 | 1 | 12 | 3 | 2 | 1 | 5 | 50 | ST-21 complex | 5 |
| 301 | OXC6845 | human | *Campylobacter jejuni* | 2 | 1 | 1 | 3 | 2 | 1 | 5 | 21 | ST-21 complex | 5 |
| 302 | OXC6850 | human | *Campylobacter jejuni* | 2 | 1 | 1 | 3 | 2 | 1 | 5 | 21 | ST-21 complex | 5 |
| 303 | OXC6852 | human | *Campylobacter jejuni* | 2 | 1 | 1 | 3 | 2 | 1 | 3 | 262 | ST-21 complex | 5 |
| 304 | OXC6854 | human | *Campylobacter jejuni* | 2 | 1 | 21 | 3 | 2 | 1 | 5 | 53 | ST-21 complex | 5 |
| 305 | OXC6856 | human | *Campylobacter jejuni* | 2 | 1 | 21 | 3 | 2 | 1 | 5 | 53 | ST-21 complex | 5 |
| 306 | OXC6857 | human | *Campylobacter jejuni* | 2 | 1 | 12 | 88 | 2 | 1 | 5 | 520 | ST-21 complex | 5 |
| 307 | OXC6860 | human | *Campylobacter jejuni* | 2 | 1 | 12 | 3 | 2 | 1 | 5 | 50 | ST-21 complex | 5 |
| 308 | OXC6861 | human | *Campylobacter jejuni* | 2 | 1 | 12 | 3 | 2 | 1 | 5 | 50 | ST-21 complex | 5 |
| 309 | OXC6862 | human | *Campylobacter jejuni* | 2 | 1 | 21 | 3 | 2 | 1 | 5 | 53 | ST-21 complex | 5 |
| 310 | OXC6865 | human | *Campylobacter jejuni* | 2 | 1 | 12 | 3 | 2 | 1 | 5 | 50 | ST-21 complex | 5 |
| 311 | OXC6867 | human | *Campylobacter jejuni* | 2 | 1 | 1 | 3 | 2 | 1 | 5 | 21 | ST-21 complex | 5 |
| 312 | OXC6868 | human | *Campylobacter jejuni* | 2 | 1 | 12 | 3 | 2 | 1 | 5 | 50 | ST-21 complex | 5 |
| 313 | OXC6872 | human | *Campylobacter jejuni* | 2 | 1 | 1 | 3 | 2 | 1 | 5 | 21 | ST-21 complex | 5 |
| 314 | OXC6876 | human | *Campylobacter jejuni* | 2 | 1 | 21 | 3 | 2 | 1 | 5 | 53 | ST-21 complex | 5 |
| 315 | OXC6879 | human | *Campylobacter jejuni* | 2 | 1 | 1 | 3 | 2 | 1 | 5 | 21 | ST-21 complex | 5 |
| 316 | OXC6886 | human | *Campylobacter jejuni* | 2 | 1 | 1 | 3 | 2 | 1 | 5 | 21 | ST-21 complex | 5 |
| 317 | OXC6892 | human | *Campylobacter jejuni* | 2 | 1 | 1 | 3 | 2 | 1 | 5 | 21 | ST-21 complex | 5 |
| 318 | OXC6900 | human | *Campylobacter jejuni* | 2 | 1 | 12 | 3 | 2 | 1 | 5 | 50 | ST-21 complex | 5 |
| 319 | OXC6909 | human | *Campylobacter jejuni* | 2 | 1 | 1 | 3 | 2 | 1 | 5 | 21 | ST-21 complex | 5 |
| 320 | OXC6922 | human | *Campylobacter jejuni* | 2 | 1 | 12 | 3 | 2 | 1 | 5 | 50 | ST-21 complex | 5 |
| 321 | OXC6924 | human | *Campylobacter jejuni* | 2 | 17 | 2 | 3 | 2 | 1 | 5 | 883 | ST-21 complex | 5 |
| 322 | OXC6927 | human | *Campylobacter jejuni* | 2 | 1 | 21 | 3 | 2 | 1 | 5 | 53 | ST-21 complex | 5 |
| 323 | OXC7002 | human | *Campylobacter jejuni* | 2 | 1 | 1 | 3 | 2 | 1 | 5 | 21 | ST-21 complex | 5 |
| 324 | OXC7015 | human | *Campylobacter jejuni* | 2 | 1 | 12 | 3 | 2 | 1 | 5 | 50 | ST-21 complex | 5 |
| 325 | OXC7020 | human | *Campylobacter jejuni* | 2 | 1 | 12 | 3 | 2 | 1 | 5 | 50 | ST-21 complex | 5 |
| 326 | OXC7029 | human | *Campylobacter jejuni* | 2 | 1 | 12 | 3 | 2 | 1 | 5 | 50 | ST-21 complex | 5 |
| 327 | OXC7031 | human | *Campylobacter jejuni* | 2 | 17 | 2 | 3 | 2 | 1 | 5 | 883 | ST-21 complex | 5 |
| 328 | OXC7035 | human | *Campylobacter jejuni* | 2 | 17 | 2 | 3 | 2 | 1 | 5 | 883 | ST-21 complex | 5 |

| | | | | | | | | | | | | |
|---|---|---|---|---|---|---|---|---|---|---|---|---|
| 329 | OXC7037 | human | *Campylobacter jejuni* | 2 | 1 | 1 | 3 | 2 | 1 | 5 | 21 | ST-21 complex | 5 |
| 330 | OXC7041 | human | *Campylobacter jejuni* | 2 | 1 | 12 | 3 | 2 | 1 | 5 | 50 | ST-21 complex | 5 |
| 331 | OXC7046 | human | *Campylobacter jejuni* | 8 | 1 | 6 | 3 | 2 | 1 | 12 | 2135 | ST-21 complex | 5 |
| 332 | OXC7049 | human | *Campylobacter jejuni* | 2 | 1 | 5 | 3 | 2 | 54 | 5 | 266 | ST-21 complex | 5 |
| 333 | OXC7058 | human | *Campylobacter jejuni* | 2 | 1 | 1 | 3 | 2 | 1 | 5 | 21 | ST-21 complex | 5 |
| 334 | OXC7065 | human | *Campylobacter jejuni* | 2 | 1 | 12 | 3 | 2 | 1 | 5 | 50 | ST-21 complex | 5 |
| 335 | OXC7068 | human | *Campylobacter jejuni* | 2 | 1 | 5 | 3 | 2 | 1 | 5 | 19 | ST-21 complex | 5 |
| 336 | OXC7071 | human | *Campylobacter jejuni* | 2 | 1 | 12 | 3 | 2 | 1 | 5 | 50 | ST-21 complex | 5 |
| 337 | OXC7073 | human | *Campylobacter jejuni* | 2 | 1 | 1 | 3 | 2 | 1 | 5 | 21 | ST-21 complex | 5 |
| 338 | OXC7078 | human | *Campylobacter jejuni* | 2 | 1 | 5 | 3 | 2 | 25 | 5 | 1949 | ST-21 complex | 5 |
| 339 | OXC7080 | human | *Campylobacter jejuni* | 2 | 1 | 12 | 3 | 2 | 1 | 5 | 50 | ST-21 complex | 5 |
| 340 | OXC7084 | human | *Campylobacter jejuni* | 2 | 1 | 1 | 3 | 2 | 1 | 5 | 21 | ST-21 complex | 5 |
| 341 | OXC7091 | human | *Campylobacter jejuni* | 2 | 1 | 1 | 3 | 2 | 1 | 3 | 262 | ST-21 complex | 5 |
| 342 | OXC7089 | human | *Campylobacter jejuni* | 2 | 1 | 21 | 3 | 2 | 1 | 5 | 53 | ST-21 complex | 5 |
| 343 | OXC7173 | human | *Campylobacter jejuni* | 8 | 1 | 6 | 3 | 2 | 1 | 1 | 44 | ST-21 complex | 5 |
| 344 | OXC7187 | human | *Campylobacter jejuni* | 2 | 1 | 1 | 3 | 2 | 1 | 5 | 21 | ST-21 complex | 5 |
| 345 | OXC7188 | human | *Campylobacter jejuni* | 2 | 1 | 12 | 3 | 2 | 1 | 5 | 50 | ST-21 complex | 5 |
| 346 | OXC7190 | human | *Campylobacter jejuni* | 2 | 1 | 1 | 3 | 2 | 1 | 5 | 21 | ST-21 complex | 5 |
| 347 | OXC7195 | human | *Campylobacter jejuni* | 2 | 1 | 1 | 3 | 2 | 1 | 5 | 21 | ST-21 complex | 5 |
| 348 | OXC6605 | human | *Campylobacter jejuni* | 2 | 1 | 1 | 3 | 2 | 1 | 5 | 21 | ST-21 complex | 5 |
| 349 | CAMP45 | chicken | *Campylobacter jejuni* | 4 | 7 | 10 | 4 | 1 | 7 | 1 | 45 | ST-45 complex | 1 |
| 350 | CampsClin11 | human | *Campylobacter jejuni* | 48 | 7 | 10 | 4 | 1 | 7 | 1 | 11 | ST-45 complex | 2 |
| 351 | chick2219 | chicken | *Campylobacter jejuni* | 10 | 7 | 10 | 4 | 1 | 7 | 1 | 2219 | ST-45 complex | 2 |
| 352 | chick594 | chicken | *Campylobacter jejuni* | 4 | 7 | 10 | 4 | 42 | 51 | 1 | 583 | ST-45 complex | 2 |
| 353 | cow334 | cattle | *Campylobacter jejuni* | 4 | 7 | 40 | 4 | 42 | 7 | 1 | 334 | ST-45 complex | 2 |
| 354 | CampsClin230 | human | *Campylobacter jejuni* | 4 | 7 | 41 | 4 | 42 | 7 | 1 | 230 | ST-45 complex | 2 |
| 355 | cowa45 | cattle | *Campylobacter jejuni* | 4 | 7 | 10 | 4 | 1 | 7 | 1 | 45 | ST-45 complex | 2 |
| 356 | chick2213 | chicken | *Campylobacter jejuni* | 4 | 7 | 40 | 4 | 42 | 7 | 1 | 334 | ST-45 complex | 2 |
| 357 | chickc45 | chicken | *Campylobacter jejuni* | 4 | 7 | 10 | 4 | 1 | 7 | 1 | 45 | ST-45 complex | 2 |
| 358 | chick11 | chicken | *Campylobacter jejuni* | 48 | 7 | 10 | 4 | 1 | 7 | 1 | 11 | ST-45 complex | 2 |

| | | | | | | | | | | | | |
|---|---|---|---|---|---|---|---|---|---|---|---|---|
| 359 | chick1003 | chicken | *Campylobacter jejuni* | 8 | 7 | 4 | 4 | 125 | 7 | 1 | 1003 | ST-45 complex | 2 |
| 360 | chick2048 | chicken | *Campylobacter jejuni* | 4 | 7 | 10 | 4 | 1 | 7 | 1 | 45 | ST-45 complex | 2 |
| 361 | chick2223 | chicken | *Campylobacter jejuni* | 4 | 7 | 10 | 4 | 1 | 7 | 1 | 45 | ST-45 complex | 2 |
| 362 | cowb45 | cattle | *Campylobacter jejuni* | 4 | 7 | 10 | 4 | 1 | 7 | 1 | 45 | ST-45 complex | 2 |
| 363 | cowc45 | cattle | *Campylobacter jejuni* | 4 | 7 | 10 | 4 | 1 | 7 | 1 | 45 | ST-45 complex | 2 |
| 364 | cowd45 | cattle | *Campylobacter jejuni* | 4 | 7 | 10 | 4 | 1 | 7 | 1 | 45 | ST-45 complex | 2 |
| 365 | cow137 | cattle | *Campylobacter jejuni* | 4 | 7 | 10 | 4 | 42 | 7 | 1 | 137 | ST-45 complex | 2 |
| 366 | cow583 | cattle | *Campylobacter jejuni* | 4 | 7 | 10 | 4 | 42 | 51 | 1 | 583 | ST-45 complex | 2 |
| 367 | cow3207 | cattle | *Campylobacter jejuni* | 4 | 7 | 40 | 4 | 42 | 7 | 1 | 334 | ST-45 complex | 3 |
| 368 | cow3214 | cattle | *Campylobacter jejuni* | 4 | 7 | 10 | 4 | 1 | 7 | 1 | 45 | ST-45 complex | 3 |
| 369 | chickb45 | chicken | *Campylobacter jejuni* | 4 | 7 | 10 | 4 | 1 | 7 | 1 | 45 | ST-45 complex | 3 |
| 370 | chickd45 | chicken | *Campylobacter jejuni* | 4 | 7 | 10 | 4 | 1 | 7 | 1 | 45 | ST-45 complex | 3 |
| 371 | chick230 | chicken | *Campylobacter jejuni* | 4 | 7 | 41 | 4 | 42 | 7 | 1 | 230 | ST-45 complex | 3 |
| 372 | OxClina45 | human | *Campylobacter jejuni* | 4 | 7 | 10 | 4 | 1 | 7 | 1 | 45 | ST-45 complex | 3 |
| 373 | starling45 | wild bird | *Campylobacter jejuni* | 4 | 7 | 10 | 4 | 1 | 7 | 1 | 45 | ST-45 complex | 3 |
| 374 | goose137 | wild bird | *Campylobacter jejuni* | 4 | 7 | 10 | 4 | 42 | 7 | 1 | 137 | ST-45 complex | 3 |
| 375 | duck45 | wild bird | *Campylobacter jejuni* | 4 | 7 | 10 | 4 | 1 | 7 | 1 | 45 | ST-45 complex | 3 |
| 376 | Cj55037 | chicken | *Campylobacter jejuni* | 4 | 7 | 10 | 4 | 1 | 7 | 1 | 45 | ST-45 complex | 4 |
| 377 | OXC5330 | human | *Campylobacter jejuni* | 8 | 7 | 4 | 4 | 125 | 7 | 1 | 1003 | ST-45 complex | 5 |
| 378 | OXC5331 | human | *Campylobacter jejuni* | 4 | 7 | 10 | 4 | 42 | 7 | 1 | 137 | ST-45 complex | 5 |
| 379 | OXC5332 | human | *Campylobacter jejuni* | 4 | 7 | 10 | 4 | 1 | 7 | 1 | 45 | ST-45 complex | 5 |
| 380 | OXC5343 | human | *Campylobacter jejuni* | 4 | 7 | 10 | 4 | 10 | 7 | 1 | 2109 | ST-45 complex | 5 |
| 381 | OXC5850 | human | *Campylobacter jejuni* | 4 | 7 | 10 | 4 | 1 | 7 | 1 | 45 | ST-45 complex | 5 |
| 382 | OXC5420 | human | *Campylobacter jejuni* | 4 | 7 | 10 | 4 | 1 | 7 | 1 | 45 | ST-45 complex | 5 |
| 383 | OXC5434 | human | *Campylobacter jejuni* | 4 | 7 | 10 | 4 | 42 | 7 | 1 | 137 | ST-45 complex | 5 |
| 384 | OXC5744 | human | *Campylobacter jejuni* | 4 | 7 | 10 | 4 | 42 | 7 | 1 | 137 | ST-45 complex | 5 |
| 385 | OXC5816 | human | *Campylobacter jejuni* | 4 | 7 | 10 | 4 | 1 | 7 | 1 | 45 | ST-45 complex | 5 |

| | | | | | | | | | | | | |
|---|---|---|---|---|---|---|---|---|---|---|---|---|
| 386 | OXC5941 | human | *Campylobacter jejuni* | 4 | 7 | 10 | 4 | 1 | 7 | 1 | 45 | ST-45 complex | 5 |
| 387 | OXC6259 | human | *Campylobacter jejuni* | 4 | 7 | 10 | 4 | 1 | 7 | 1 | 45 | ST-45 complex | 5 |
| 388 | OXC6278 | human | *Campylobacter jejuni* | 4 | 7 | 10 | 4 | 1 | 7 | 1 | 45 | ST-45 complex | 5 |
| 389 | OXC6293 | human | *Campylobacter jejuni* | 4 | 7 | 10 | 4 | 1 | 7 | 1 | 45 | ST-45 complex | 5 |
| 390 | OXC6294 | human | *Campylobacter jejuni* | 4 | 7 | 10 | 4 | 42 | 51 | 1 | 583 | ST-45 complex | 5 |
| 391 | OXC6313 | human | *Campylobacter jejuni* | 4 | 7 | 10 | 4 | 1 | 7 | 1 | 45 | ST-45 complex | 5 |
| 392 | OXC6314 | human | *Campylobacter jejuni* | 4 | 7 | 10 | 4 | 1 | 7 | 1 | 45 | ST-45 complex | 5 |
| 393 | OXC6321 | human | *Campylobacter jejuni* | 4 | 7 | 10 | 4 | 1 | 7 | 1 | 45 | ST-45 complex | 5 |
| 394 | OXC6330 | human | *Campylobacter jejuni* | 4 | 7 | 10 | 4 | 42 | 51 | 1 | 583 | ST-45 complex | 5 |
| 395 | OXC6339 | human | *Campylobacter jejuni* | 4 | 7 | 10 | 4 | 1 | 7 | 1 | 45 | ST-45 complex | 5 |
| 396 | OXC6351 | human | *Campylobacter jejuni* | 4 | 7 | 10 | 4 | 1 | 7 | 1 | 45 | ST-45 complex | 5 |
| 397 | OXC6358 | human | *Campylobacter jejuni* | 2 | 7 | 10 | 4 | 1 | 7 | 1 | 233 | ST-45 complex | 5 |
| 398 | OXC6365 | human | *Campylobacter jejuni* | 4 | 7 | 10 | 4 | 1 | 7 | 1 | 45 | ST-45 complex | 5 |
| 399 | OXC6419 | human | *Campylobacter jejuni* | 4 | 7 | 10 | 4 | 1 | 7 | 1 | 45 | ST-45 complex | 5 |
| 400 | OXC6437 | human | *Campylobacter jejuni* | 4 | 7 | 10 | 4 | 1 | 7 | 1 | 45 | ST-45 complex | 5 |
| 401 | OXC6448 | human | *Campylobacter jejuni* | 4 | 7 | 10 | 4 | 42 | 7 | 1 | 137 | ST-45 complex | 5 |
| 402 | OXC6515 | human | *Campylobacter jejuni* | 4 | 7 | 73 | 4 | 1 | 7 | 1 | 845 | ST-45 complex | 5 |
| 403 | OXC6536 | human | *Campylobacter jejuni* | 4 | 7 | 10 | 4 | 1 | 7 | 1 | 45 | ST-45 complex | 5 |
| 404 | OXC6592 | human | *Campylobacter jejuni* | 4 | 7 | 10 | 4 | 42 | 7 | 1 | 137 | ST-45 complex | 5 |
| 405 | OXC6614 | human | *Campylobacter jejuni* | 2 | 7 | 10 | 4 | 1 | 7 | 1 | 233 | ST-45 complex | 5 |
| 406 | OXC6624 | human | *Campylobacter jejuni* | 4 | 7 | 10 | 4 | 1 | 7 | 1 | 45 | ST-45 complex | 5 |
| 407 | OXC6797 | human | *Campylobacter jejuni* | 4 | 7 | 10 | 4 | 1 | 7 | 1 | 45 | ST-45 complex | 5 |
| 408 | OXC6803 | human | *Campylobacter jejuni* | 4 | 7 | 10 | 4 | 1 | 7 | 1 | 45 | ST-45 complex | 5 |
| 409 | OXC6819 | human | *Campylobacter jejuni* | 4 | 7 | 10 | 4 | 42 | 51 | 1 | 583 | ST-45 complex | 5 |
| 410 | OXC6938 | human | *Campylobacter jejuni* | 4 | 7 | 10 | 4 | 10 | 7 | 1 | 2109 | ST-45 complex | 5 |
| 411 | OXC6970 | human | *Campylobacter jejuni* | 4 | 7 | 10 | 4 | 42 | 51 | 1 | 583 | ST-45 complex | 5 |
| 412 | OXC7103 | human | *Campylobacter jejuni* | 4 | 7 | 10 | 4 | 1 | 7 | 1 | 45 | ST-45 complex | 5 |
| 413 | OXC7113 | human | *Campylobacter jejuni* | 4 | 7 | 10 | 4 | 42 | 25 | 1 | 538 | ST-45 complex | 5 |
| 414 | OXC7129 | human | *Campylobacter jejuni* | 4 | 7 | 10 | 4 | 1 | 7 | 1 | 45 | ST-45 complex | 5 |
| 415 | OXC7167 | human | *Campylobacter jejuni* | 4 | 7 | 10 | 4 | 1 | 532 | 1 | 6886 | ST-45 complex | 5 |
| 416 | OXC7169 | human | *Campylobacter jejuni* | 4 | 7 | 10 | 4 | 1 | 7 | 1 | 45 | ST-45 complex | 5 |
| 417 | OXC6894 | human | *Campylobacter jejuni* | 4 | 7 | 10 | 4 | 1 | 7 | 1 | 45 | ST-45 complex | 5 |
| 418 | OXC6895 | human | *Campylobacter jejuni* | 4 | 7 | 10 | 4 | 1 | 7 | 1 | 45 | ST-45 complex | 5 |
| 419 | OXC6896 | human | *Campylobacter jejuni* | 4 | 7 | 10 | 4 | 1 | 7 | 1 | 45 | ST-45 complex | 5 |
| 420 | OXC6916 | human | *Campylobacter jejuni* | 4 | 7 | 10 | 4 | 1 | 7 | 1 | 45 | ST-45 complex | 5 |
| 421 | OXC7014 | human | *Campylobacter jejuni* | 4 | 7 | 10 | 4 | 1 | 7 | 1 | 45 | ST-45 complex | 5 |
| 422 | OXC7018 | human | *Campylobacter jejuni* | 4 | 7 | 10 | 4 | 1 | 7 | 1 | 45 | ST-45 complex | 5 |
| 423 | OXC7025 | human | *Campylobacter jejuni* | 4 | 7 | 10 | 4 | 1 | 7 | 1 | 45 | ST-45 complex | 5 |
| 424 | OXC7026 | human | *Campylobacter jejuni* | 10 | 7 | 10 | 4 | 1 | 7 | 1 | 2219 | ST-45 complex | 5 |
| 425 | OXC7032 | human | *Campylobacter jejuni* | 4 | 7 | 10 | 4 | 42 | 51 | 1 | 583 | ST-45 complex | 5 |
| 426 | OXC7061 | human | *Campylobacter jejuni* | 4 | 7 | 10 | 4 | 1 | 7 | 1 | 45 | ST-45 complex | 5 |
| 427 | OXC7067 | human | *Campylobacter jejuni* | 4 | 7 | 10 | 4 | 42 | 7 | 1 | 137 | ST-45 complex | 5 |
| 428 | OXC7074 | human | *Campylobacter jejuni* | 4 | 7 | 10 | 4 | 1 | 7 | 1 | 45 | ST-45 complex | 5 |
| 429 | OXC7090 | human | *Campylobacter jejuni* | 4 | 7 | 10 | 4 | 1 | 7 | 1 | 45 | ST-45 complex | 5 |
| 430 | OXC7093 | human | *Campylobacter jejuni* | 48 | 7 | 10 | 4 | 1 | 7 | 1 | 11 | ST-45 complex | 5 |
| 431 | OXC7099 | human | *Campylobacter jejuni* | 4 | 7 | 10 | 4 | 1 | 7 | 1 | 45 | ST-45 complex | 5 |
| 432 | OXC7170 | human | *Campylobacter jejuni* | 4 | 7 | 10 | 4 | 42 | 7 | 1 | 137 | ST-45 complex | 5 |
| 433 | OXC7179 | human | *Campylobacter jejuni* | 4 | 7 | 10 | 4 | 1 | 7 | 1 | 45 | ST-45 complex | 5 |
| 434 | OXC7185 | human | *Campylobacter jejuni* | 4 | 7 | 10 | 4 | 1 | 7 | 1 | 45 | ST-45 complex | 5 |
| 435 | OXC7186 | human | *Campylobacter jejuni* | 4 | 7 | 10 | 4 | 1 | 7 | 1 | 45 | ST-45 complex | 5 |
| 436 | 171 | human | *Campylobacter coli* | 33 | 38 | 30 | 115 | 104 | 85 | 17 | 867 | ST-828 complex | 1 |
| 437 | CAMP886 | pig | *Campylobacter coli* | 33 | 38 | 30 | 82 | 104 | 85 | 68 | 887 | ST-828 complex | 1 |
| 438 | CAMP2588 | chicken | *Campylobacter coli* | 33 | 38 | 30 | 115 | 104 | 85 | 17 | 867 | ST-828 complex | 1 |
| 439 | CAMP3129 | human | *Campylobacter coli* | 157 | 39 | 30 | 79 | 668 | 35 | 17 | | ST-828 complex | 1 |
| 440 | CAMP1090 | chicken | *Campylobacter coli* | 33 | 39 | 30 | 82 | 104 | 43 | 17 | 828 | ST-828 complex | 2 |
| 441 | CAMP828 | chicken | *Campylobacter coli* | 33 | 39 | 30 | 82 | 104 | 43 | 17 | 828 | ST-828 complex | 2 |
| 442 | cow3202 | cattle | *Campylobacter coli* | 33 | 39 | 30 | 82 | 104 | 56 | 17 | 827 | ST-828 complex | 2 |
| 443 | Cc111-3 | pig | *Campylobacter coli* | 33 | 38 | 30 | 167 | 104 | 43 | 17 | 1467 | ST-828 complex | 4 |
| 444 | CcZ163 | chicken | *Campylobacter coli* | 33 | 38 | 30 | 82 | 104 | 332 | 17 | 3336 | ST-828 complex | 4 |
| 445 | Cc2553 | chicken | *Campylobacter coli* | 33 | 39 | 30 | 82 | 113 | 47 | 17 | 825 | ST-828 complex | 4 |

| | | | | | | | | | | | | |
|---|---|---|---|---|---|---|---|---|---|---|---|---|
| 446 | Cc2680 | chicken | *Campylobacter coli* | 33 | 39 | 103 | 82 | 104 | 324 | 41 | 3872 | ST-828 complex | 4 |
| 447 | Cc2685 | chicken | *Campylobacter coli* | 33 | 39 | 30 | 82 | 211 | 85 | 17 | 1082 | ST-828 complex | 4 |
| 448 | Cc2688 | chicken | *Campylobacter coli* | 33 | 39 | 30 | 82 | 104 | 43 | 41 | 1017 | ST-828 complex | 4 |
| 449 | Cc2692 | chicken | *Campylobacter coli* | 33 | 39 | 30 | 79 | 113 | 47 | 17 | 860 | ST-828 complex | 4 |
| 450 | Cc2698 | chicken | *Campylobacter coli* | 33 | 39 | 30 | 82 | 113 | 43 | 17 | 829 | ST-828 complex | 4 |
| 451 | Cc84-2 | pig | *Campylobacter coli* | 33 | 38 | 30 | 78 | 104 | 35 | 17 | 1113 | ST-828 complex | 4 |
| 452 | Cc80352 | chicken | *Campylobacter coli* | 33 | 39 | 30 | 82 | 104 | 43 | 41 | 1017 | ST-828 complex | 4 |
| 453 | Cc86119 | chicken | *Campylobacter coli* | 33 | 39 | 30 | 82 | 113 | 47 | 17 | 825 | ST-828 complex | 4 |
| 454 | Cc1091 | cattle | *Campylobacter coli* | 33 | 39 | 30 | 78 | 104 | 43 | 17 | 1068 | ST-828 complex | 4 |
| 455 | Cc1098 | cattle | *Campylobacter coli* | 33 | 39 | 30 | 82 | 104 | 85 | 68 | 1104 | ST-828 complex | 4 |
| 456 | Cc1148 | cattle | *Campylobacter coli* | 33 | 39 | 30 | 78 | 104 | 43 | 17 | 1068 | ST-828 complex | 4 |
| 457 | Cc1417 | cattle | *Campylobacter coli* | 33 | 153 | 44 | 82 | 104 | 44 | 17 | 3221 | ST-828 complex | 4 |
| 458 | Cc132-6 | pig | *Campylobacter coli* | 33 | 153 | 44 | 82 | 104 | 44 | 17 | 3221 | ST-828 complex | 4 |
| 459 | Cc1891 | cattle | *Campylobacter coli* | 33 | 39 | 30 | 78 | 104 | 43 | 17 | 1068 | ST-828 complex | 4 |
| 460 | Cc1909 | cattle | *Campylobacter coli* | 33 | 39 | 30 | 82 | 104 | 85 | 68 | 1104 | ST-828 complex | 4 |
| 461 | Cc59-2 | pig | *Campylobacter coli* | 33 | 38 | 30 | 82 | 104 | 35 | 36 | 890 | ST-828 complex | 4 |
| 462 | Cc1948 | cattle | *Campylobacter coli* | 33 | 39 | 30 | 82 | 104 | 85 | 68 | 1104 | ST-828 complex | 4 |
| 463 | Cc1957 | cattle | *Campylobacter coli* | 33 | 153 | 30 | 82 | 104 | 85 | 68 | 2698 | ST-828 complex | 4 |
| 464 | Cc1961 | cattle | *Campylobacter coli* | 33 | 39 | 30 | 82 | 104 | 85 | 68 | 1104 | ST-828 complex | 4 |
| 465 | Cc202/04 | human | *Campylobacter coli* | 33 | 39 | 30 | 82 | 189 | 43 | 17 | 1585 | ST-828 complex | 4 |
| 466 | Cc67-8 | pig | *Campylobacter coli* | 32 | 39 | 44 | 82 | 104 | 43 | 36 | 1061 | ST-828 complex | 4 |
| 467 | CcLMG9854 | human | *Campylobacter coli* | 33 | 39 | 30 | 78 | 104 | 43 | 17 | 1068 | ST-828 complex | 4 |
| 468 | CcLMG23336 | human | *Campylobacter coli* | 33 | 176 | 30 | 82 | 451 | 43 | 17 | 3868 | ST-828 complex | 4 |
| 469 | CcLMG23341 | human | *Campylobacter coli* | 33 | 39 | 30 | 79 | 104 | 35 | 17 | 855 | ST-828 complex | 4 |
| 470 | CcLMG23342 | human | *Campylobacter coli* | 33 | 39 | 30 | 79 | 104 | 35 | 17 | 855 | ST-828 complex | 4 |
| 471 | CcLMG23344 | human | *Campylobacter coli* | 33 | 176 | 30 | 82 | 113 | 43 | 17 | 1586 | ST-828 complex | 4 |
| 472 | CcLMG9853 | human | *Campylobacter coli* | 33 | 39 | 261 | 79 | 104 | 64 | 17 | 3869 | ST-828 complex | 4 |
| 473 | CcH6 | human | *Campylobacter coli* | 33 | 176 | 30 | 79 | 113 | 43 | 17 | 3020 | ST-828 complex | 4 |
| 474 | CcH8 | human | *Campylobacter coli* | 33 | 39 | 30 | 79 | 104 | 43 | 41 | 901 | ST-828 complex | 4 |
| 475 | CCH9 | human | *Campylobacter coli* | 33 | 39 | 30 | 82 | 113 | 47 | 17 | 825 | ST-828 complex | 4 |

| | | | | | | | | | | | | |
|---|---|---|---|---|---|---|---|---|---|---|---|---|
| 476 | CcH56 | human | *Campylobacter coli* | 33 | 38 | 30 | 82 | 104 | 35 | 17 | 1096 | ST-828 complex | 4 |
| 477 | CcZ156 | chicken | *Campylobacter coli* | 33 | 38 | 30 | 82 | 104 | 43 | 17 | 854 | ST-828 complex | 4 |
| 478 | OXC5348 | human | *Campylobacter coli* | 33 | 39 | 30 | 82 | 104 | 56 | 17 | 827 | ST-828 complex | 5 |
| 479 | OXC5353 | human | *Campylobacter coli* | 33 | 39 | 30 | 79 | 530 | 43 | 17 | 5149 | ST-828 complex | 5 |
| 480 | OXC5363 | human | *Campylobacter coli* | 33 | 39 | 30 | 82 | 104 | 56 | 17 | 827 | ST-828 complex | 5 |
| 481 | OXC5370 | human | *Campylobacter coli* | 33 | 39 | 30 | 82 | 104 | 56 | 17 | 827 | ST-828 complex | 5 |
| 482 | OXC5386 | human | *Campylobacter coli* | 33 | 39 | 30 | 82 | 113 | 47 | 17 | 825 | ST-828 complex | 5 |
| 483 | OXC5677 | human | *Campylobacter coli* | 33 | 66 | 30 | 174 | 188 | 43 | 17 | 6132 | ST-828 complex | 5 |
| 484 | OXC5685 | human | *Campylobacter coli* | 33 | 39 | 30 | 82 | 112 | 56 | 17 | 1578 | ST-828 complex | 5 |
| 485 | OXC5688 | human | *Campylobacter coli* | 33 | 39 | 30 | 82 | 104 | 56 | 17 | 827 | ST-828 complex | 5 |
| 486 | OXC5705 | human | *Campylobacter coli* | 33 | 39 | 30 | 82 | 113 | 44 | 17 | 872 | ST-828 complex | 5 |
| 487 | OXC5482 | human | *Campylobacter coli* | 33 | 39 | 65 | 79 | 104 | 85 | 17 | 2273 | ST-828 complex | 5 |
| 488 | OXC5605 | human | *Campylobacter coli* | 33 | 39 | 30 | 79 | 113 | 43 | 17 | 832 | ST-828 complex | 5 |
| 489 | OXC5629 | human | *Campylobacter coli* | 33 | 39 | 30 | 79 | 104 | 35 | 17 | 855 | ST-828 complex | 5 |
| 490 | OXC5635 | human | *Campylobacter coli* | 124 | 39 | 30 | 79 | 104 | 47 | 17 | 1541 | ST-828 complex | 5 |
| 491 | OXC5837 | human | *Campylobacter coli* | 33 | 39 | 30 | 82 | 104 | 56 | 17 | 827 | ST-828 complex | 5 |
| 492 | OXC5842 | human | *Campylobacter coli* | 33 | 39 | 66 | 82 | 104 | 44 | 174 | 2178 | ST-828 complex | 5 |
| 493 | OXC5849 | human | *Campylobacter coli* | 33 | 39 | 30 | 79 | 104 | 47 | 17 | 830 | ST-828 complex | 5 |
| 494 | OXC5853 | human | *Campylobacter coli* | 33 | 39 | 30 | 79 | 104 | 47 | 17 | 830 | ST-828 complex | 5 |
| 495 | OXC5855 | human | *Campylobacter coli* | 33 | 39 | 30 | 82 | 113 | 47 | 17 | 825 | ST-828 complex | 5 |
| 496 | OXC5856 | human | *Campylobacter coli* | 33 | 39 | 30 | 79 | 104 | 47 | 17 | 830 | ST-828 complex | 5 |
| 497 | OXC5864 | human | *Campylobacter coli* | 33 | 39 | 30 | 82 | 189 | 47 | 17 | 1191 | ST-828 complex | 5 |
| 498 | OXC5473 | human | *Campylobacter coli* | 33 | 39 | 30 | 82 | 23 | 43 | 17 | 6131 | ST-828 complex | 5 |
| 499 | OXC5723 | human | *Campylobacter coli* | 33 | 39 | 30 | 82 | 104 | 56 | 17 | 827 | ST-828 complex | 5 |
| 500 | OXC5742 | human | *Campylobacter coli* | 33 | 39 | 30 | 82 | 113 | 44 | 17 | 872 | ST-828 complex | 5 |
| 501 | OXC5763 | human | *Campylobacter coli* | 33 | 39 | 30 | 82 | 113 | 47 | 17 | 825 | ST-828 complex | 5 |
| 502 | OXC5773 | human | *Campylobacter coli* | 33 | 39 | 30 | 82 | 104 | 43 | 17 | 828 | ST-828 complex | 5 |
| 503 | OXC5796 | human | *Campylobacter coli* | 33 | 39 | 30 | 82 | 113 | 44 | 17 | 872 | ST-828 complex | 5 |
| 504 | OXC5810 | human | *Campylobacter coli* | 33 | 39 | 30 | 82 | 189 | 44 | 17 | 5165 | ST-828 complex | 5 |
| 505 | OXC5827 | human | *Campylobacter coli* | 33 | 39 | 30 | 79 | 113 | 47 | 17 | 860 | ST-828 complex | 5 |

| | | | | | | | | | | | | |
|---|---|---|---|---|---|---|---|---|---|---|---|---|
| 506 | OXC5831 | human | *Campylobacter coli* | 33 | 39 | 30 | 82 | 189 | 47 | 17 | 1191 | ST-828 complex | 5 |
| 507 | OXC5922 | human | *Campylobacter coli* | 33 | 39 | 30 | 82 | 104 | 56 | 17 | 827 | ST-828 complex | 5 |
| 508 | OXC5923 | human | *Campylobacter coli* | 33 | 39 | 30 | 82 | 104 | 56 | 17 | 827 | ST-828 complex | 5 |
| 509 | OXC6559 | human | *Campylobacter coli* | 33 | 38 | 30 | 466 | 104 | 43 | 17 | 5757 | ST-828 complex | 5 |
| 510 | OXC6568 | human | *Campylobacter coli* | 33 | 39 | 30 | 82 | 113 | 44 | 17 | 872 | ST-828 complex | 5 |
| 511 | OXC6576 | human | *Campylobacter coli* | 33 | 39 | 30 | 79 | 104 | 35 | 17 | 855 | ST-828 complex | 5 |
| 512 | OXC6577 | human | *Campylobacter coli* | 33 | 39 | 30 | 82 | 113 | 44 | 17 | 872 | ST-828 complex | 5 |
| 513 | OXC6253 | human | *Campylobacter coli* | 33 | 39 | 30 | 82 | 113 | 43 | 17 | 829 | ST-828 complex | 5 |
| 514 | OXC6258 | human | *Campylobacter coli* | 33 | 39 | 30 | 79 | 113 | 47 | 17 | 860 | ST-828 complex | 5 |
| 515 | OXC6263 | human | *Campylobacter coli* | 33 | 39 | 30 | 79 | 104 | 47 | 17 | 830 | ST-828 complex | 5 |
| 516 | OXC6267 | human | *Campylobacter coli* | 33 | 39 | 30 | 79 | 247 | 43 | 17 | 5642 | ST-828 complex | 5 |
| 517 | OXC6276 | human | *Campylobacter coli* | 33 | 39 | 65 | 82 | 596 | 47 | 17 | 5733 | ST-828 complex | 5 |
| 518 | OXC6296 | human | *Campylobacter coli* | 87 | 39 | 30 | 79 | 243 | 43 | 17 | 5737 | ST-828 complex | 5 |
| 519 | OXC6297 | human | *Campylobacter coli* | 33 | 39 | 30 | 82 | 113 | 47 | 17 | 825 | ST-828 complex | 5 |
| 520 | OXC6308 | human | *Campylobacter coli* | 33 | 39 | 30 | 82 | 104 | 56 | 17 | 827 | ST-828 complex | 5 |
| 521 | OXC6309 | human | *Campylobacter coli* | 33 | 110 | 30 | 115 | 104 | 43 | 17 | 5734 | ST-828 complex | 5 |
| 522 | OXC6312 | human | *Campylobacter coli* | 33 | 39 | 30 | 82 | 104 | 56 | 17 | 827 | ST-828 complex | 5 |
| 523 | OXC6337 | human | *Campylobacter coli* | 33 | 39 | 65 | 140 | 104 | 43 | 17 | 1600 | ST-828 complex | 5 |
| 524 | OXC6338 | human | *Campylobacter coli* | 33 | 39 | 30 | 79 | 104 | 206 | 17 | 1628 | ST-828 complex | 5 |
| 525 | OXC6343 | human | *Campylobacter coli* | 33 | 39 | 30 | 82 | 104 | 47 | 36 | 962 | ST-828 complex | 5 |
| 526 | OXC6371 | human | *Campylobacter coli* | 33 | 110 | 30 | 115 | 104 | 43 | 17 | 5734 | ST-828 complex | 5 |
| 527 | OXC6372 | human | *Campylobacter coli* | 33 | 39 | 30 | 139 | 113 | 47 | 17 | 4425 | ST-828 complex | 5 |
| 528 | OXC6376 | human | *Campylobacter coli* | 33 | 110 | 30 | 115 | 104 | 43 | 17 | 5734 | ST-828 complex | 5 |
| 529 | OXC6378 | human | *Campylobacter coli* | 33 | 39 | 30 | 139 | 113 | 47 | 17 | 4425 | ST-828 complex | 5 |
| 530 | OXC6380 | human | *Campylobacter coli* | 33 | 39 | 30 | 82 | 113 | 47 | 353 | 5735 | ST-828 complex | 5 |
| 531 | OXC6385 | human | *Campylobacter coli* | 33 | 39 | 30 | 79 | 104 | 206 | 17 | 1628 | ST-828 complex | 5 |
| 532 | OXC6386 | human | *Campylobacter coli* | 33 | 39 | 30 | 82 | 113 | 43 | 17 | 829 | ST-828 complex | 5 |
| 533 | OXC6400 | human | *Campylobacter coli* | 33 | 39 | 30 | 82 | 113 | 47 | 17 | 825 | ST-828 complex | 5 |
| 534 | OXC6416 | human | *Campylobacter coli* | 33 | 39 | 30 | 82 | 104 | 44 | 17 | 1145 | ST-828 complex | 5 |
| 535 | OXC6424 | human | *Campylobacter coli* | 33 | 39 | 30 | 82 | 113 | 47 | 17 | 825 | ST-828 complex | 5 |

| | | | | | | | | | | | | |
|---|---|---|---|---|---|---|---|---|---|---|---|---|
| 536 | OXC6426 | human | *Campylobacter coli* | 32 | 66 | 66 | 82 | 104 | 43 | 17 | 2464 | ST-828 complex | 5 |
| 537 | OXC6428 | human | *Campylobacter coli* | 33 | 39 | 30 | 79 | 113 | 43 | 17 | 832 | ST-828 complex | 5 |
| 538 | OXC6447 | human | *Campylobacter coli* | 33 | 39 | 30 | 82 | 104 | 56 | 17 | 827 | ST-828 complex | 5 |
| 539 | OXC6460 | human | *Campylobacter coli* | 124 | 39 | 30 | 79 | 104 | 47 | 17 | 1541 | ST-828 complex | 5 |
| 540 | OXC6471 | human | *Campylobacter coli* | 33 | 39 | 30 | 82 | 104 | 56 | 17 | 827 | ST-828 complex | 5 |
| 541 | OXC6472 | human | *Campylobacter coli* | 33 | 39 | 30 | 82 | 104 | 56 | 17 | 827 | ST-828 complex | 5 |
| 542 | OXC6474 | human | *Campylobacter coli* | 33 | 39 | 30 | 82 | 104 | 56 | 17 | 827 | ST-828 complex | 5 |
| 543 | OXC6476 | human | *Campylobacter coli* | 33 | 39 | 30 | 79 | 113 | 47 | 17 | 860 | ST-828 complex | 5 |
| 544 | OXC6504 | human | *Campylobacter coli* | 33 | 39 | 30 | 79 | 113 | 43 | 17 | 832 | ST-828 complex | 5 |
| 545 | OXC6513 | human | *Campylobacter coli* | 33 | 39 | 30 | 82 | 104 | 43 | 17 | 828 | ST-828 complex | 5 |
| 546 | OXC6523 | human | *Campylobacter coli* | 33 | 66 | 30 | 82 | 113 | 206 | 17 | 5659 | ST-828 complex | 5 |
| 547 | OXC6537 | human | *Campylobacter coli* | 124 | 39 | 30 | 79 | 104 | 47 | 17 | 1541 | ST-828 complex | 5 |
| 548 | OXC6597 | human | *Campylobacter coli* | 33 | 39 | 30 | 79 | 104 | 35 | 17 | 855 | ST-828 complex | 5 |
| 549 | OXC6601 | human | *Campylobacter coli* | 33 | 39 | 30 | 79 | 104 | 85 | 17 | 1614 | ST-828 complex | 5 |
| 550 | OXC6630 | human | *Campylobacter coli* | 33 | 39 | 30 | 82 | 104 | 44 | 17 | 1145 | ST-828 complex | 5 |
| 551 | OXC6651 | human | *Campylobacter coli* | 33 | 39 | 66 | 174 | 104 | 43 | 41 | 1181 | ST-828 complex | 5 |
| 552 | OXC6684 | human | *Campylobacter coli* | 33 | 39 | 30 | 82 | 104 | 47 | 36 | 962 | ST-828 complex | 5 |
| 553 | OXC6685 | human | *Campylobacter coli* | 33 | 39 | 30 | 82 | 104 | 56 | 17 | 827 | ST-828 complex | 5 |
| 554 | OXC6705 | human | *Campylobacter coli* | 33 | 39 | 65 | 79 | 104 | 35 | 17 | 1957 | ST-828 complex | 5 |
| 555 | OXC6710 | human | *Campylobacter coli* | 2 | 39 | 30 | 82 | 113 | 47 | 17 | 5810 | ST-828 complex | 5 |
| 556 | OXC6725 | human | *Campylobacter coli* | 33 | 38 | 30 | 82 | 104 | 35 | 17 | 1096 | ST-828 complex | 5 |
| 557 | OXC6735 | human | *Campylobacter coli* | 33 | 39 | 30 | 82 | 104 | 44 | 17 | 1145 | ST-828 complex | 5 |
| 558 | OXC6738 | human | *Campylobacter coli* | 33 | 39 | 65 | 79 | 104 | 85 | 17 | 2273 | ST-828 complex | 5 |
| 559 | OXC6744 | human | *Campylobacter coli* | 53 | 39 | 30 | 82 | 603 | 43 | 17 | 5813 | ST-828 complex | 5 |
| 560 | OXC6761 | human | *Campylobacter coli* | 33 | 39 | 30 | 82 | 104 | 56 | 17 | 827 | ST-828 complex | 5 |
| 561 | OXC6765 | | *Campylobacter coli* | 33 | 39 | 30 | 79 | 113 | 43 | 17 | 832 | ST-828 complex | 5 |
| 562 | OXC6785 | human | *Campylobacter coli* | 33 | 39 | 30 | 82 | 104 | 56 | 17 | 827 | ST-828 complex | 5 |
| 563 | OXC6810 | human | *Campylobacter coli* | 33 | 39 | 30 | 82 | 113 | 47 | 17 | 825 | ST-828 complex | 5 |
| 564 | OXC6817 | human | *Campylobacter coli* | 33 | 39 | 30 | 82 | 104 | 47 | 36 | 962 | ST-828 complex | 5 |
| 565 | OXC6825 | human | *Campylobacter coli* | 33 | 39 | 30 | 79 | 113 | 43 | 41 | 2183 | ST-828 complex | 5 |

| | | | | | | | | | | | | |
|---|---|---|---|---|---|---|---|---|---|---|---|---|
| 566 | OXC6933 | human | *Campylobacter coli* | 33 | 39 | 30 | 82 | 104 | 56 | 17 | 827 | ST-828 complex | 5 |
| 567 | OXC6937 | human | *Campylobacter coli* | 33 | 176 | 30 | 82 | 104 | 43 | 17 | 3753 | ST-828 complex | 5 |
| 568 | OXC6962 | human | *Campylobacter coli* | 33 | 39 | 65 | 79 | 104 | 85 | 17 | 2273 | ST-828 complex | 5 |
| 569 | OXC6984 | human | *Campylobacter coli* | 33 | 39 | 30 | 82 | 104 | 47 | 17 | 1055 | ST-828 complex | 5 |
| 570 | OXC6987 | human | *Campylobacter coli* | 33 | 39 | 30 | 82 | 113 | 47 | 17 | 825 | ST-828 complex | 5 |
| 571 | OXC6996 | human | *Campylobacter coli* | 33 | 39 | 30 | 82 | 104 | 56 | 17 | 827 | ST-828 complex | 5 |
| 572 | OXC7102 | human | *Campylobacter coli* | 33 | 39 | 30 | 82 | 189 | 47 | 17 | 1191 | ST-828 complex | 5 |
| 573 | OXC7110 | human | *Campylobacter coli* | 33 | 39 | 30 | 82 | 189 | 47 | 17 | 1191 | ST-828 complex | 5 |
| 574 | OXC7124 | human | *Campylobacter coli* | 33 | 39 | 30 | 82 | 104 | 56 | 17 | 827 | ST-828 complex | 5 |
| 575 | OXC7130 | human | *Campylobacter coli* | 33 | 39 | 30 | 79 | 113 | 43 | 41 | 2183 | ST-828 complex | 5 |
| 576 | OXC7131 | human | *Campylobacter coli* | 33 | 39 | 30 | 82 | 104 | 43 | 17 | 828 | ST-828 complex | 5 |
| 577 | OXC7154 | human | *Campylobacter coli* | 33 | 39 | 30 | 82 | 104 | 43 | 17 | 828 | ST-828 complex | 5 |
| 578 | OXC7164 | human | *Campylobacter coli* | 33 | 39 | 30 | 82 | 104 | 56 | 17 | 827 | ST-828 complex | 5 |
| 579 | OXC6847 | human | *Campylobacter coli* | 33 | 39 | 30 | 82 | 113 | 43 | 17 | 829 | ST-828 complex | 5 |
| 580 | OXC6864 | human | *Campylobacter coli* | 33 | 39 | 30 | 82 | 113 | 43 | 17 | 829 | ST-828 complex | 5 |
| 581 | OXC6873 | human | *Campylobacter coli* | 33 | 39 | 30 | 81 | 561 | 47 | 17 | 5349 | ST-828 complex | 5 |
| 582 | OXC6901 | human | *Campylobacter coli* | 33 | 39 | 30 | 82 | 104 | 56 | 17 | 827 | ST-828 complex | 5 |
| 583 | OXC6920 | human | *Campylobacter coli* | 33 | 39 | 30 | 82 | 104 | 56 | 17 | 827 | ST-828 complex | 5 |
| 584 | OXC7027 | human | *Campylobacter coli* | 33 | 38 | 30 | 82 | 118 | 43 | 17 | 1016 | ST-828 complex | 5 |
| 585 | OXC7051 | human | *Campylobacter coli* | 33 | 38 | 30 | 82 | 118 | 43 | 17 | 1016 | ST-828 complex | 5 |
| 586 | OXC7054 | human | *Campylobacter coli* | 33 | 39 | 30 | 82 | 113 | 47 | 17 | 825 | ST-828 complex | 5 |
| 587 | OXC7070 | human | *Campylobacter coli* | 33 | 39 | 30 | 82 | 104 | 56 | 17 | 827 | ST-828 complex | 5 |
| 588 | OXC7082 | human | *Campylobacter coli* | 33 | 39 | 30 | 82 | 113 | 47 | 17 | 825 | ST-828 complex | 5 |
| 589 | OXC7083 | human | *Campylobacter coli* | 33 | 39 | 30 | 79 | 104 | 35 | 17 | 855 | ST-828 complex | 5 |
| 590 | OXC7097 | human | *Campylobacter coli* | 33 | 39 | 30 | 82 | 104 | 56 | 17 | 827 | ST-828 complex | 5 |
| 591 | OXC7177 | human | *Campylobacter coli* | 33 | 39 | 30 | 139 | 113 | 47 | 17 | 4425 | ST-828 complex | 5 |
| 592 | OXC7199 | human | *Campylobacter coli* | 33 | 39 | 30 | 82 | 104 | 56 | 17 | 827 | ST-828 complex | 5 |
| 593 | OXC7200 | human | *Campylobacter coli* | 33 | 39 | 30 | 82 | 113 | 43 | 17 | 829 | ST-828 complex | 5 |